\documentclass[journal]{IEEEtran}

\usepackage{amsfonts}
\usepackage{amssymb}
\usepackage{cite}
\usepackage[cmex10]{amsmath}
\usepackage{float}
\usepackage{color}
\usepackage{stfloats}
\usepackage{mathrsfs}
\usepackage{algorithm}
\usepackage{algorithmic}
\usepackage{amsthm}
\usepackage{multirow}
\usepackage{changepage}
\usepackage[normalem]{ulem}
\usepackage{caption}
\usepackage{booktabs}

\newtheorem{lemma}{Lemma}
\newtheorem{proposition}{Proposition}

\newtheorem{remark}{Remark}

\IEEEoverridecommandlockouts

\ifCLASSINFOpdf
\usepackage[pdftex]{graphicx}
\DeclareGraphicsExtensions{.pdf,.jpeg,.png}
\else
\usepackage[dvips]{graphicx}
\DeclareGraphicsExtensions{.eps}
\fi
\usepackage{subfigure}
\usepackage{fancybox,dashbox}
\begin{document}
\title{Antenna Selection in MIMO Non-Orthogonal Multiple Access Systems}

\author{Yuehua~Yu,~He~Chen,~Yonghui~Li,~Zhiguo Ding, Lingyang Song,~and Branka Vucetic

\thanks{Y.~Yu,~H.~Chen,~Y.~Li,~and B.~Vucetic are with the School
of Electrical and Information Engineering, The University of Sydney, Sydney,
NSW 2006, Australia (email: \{yuehua.yu,~he.chen,~yonghui.li,~branka.vucetic\}@sydney.edu.au).}
\thanks{Z.~Ding is with the School of Computing and Communications, Lancaster University, Lancaster LA1 4YW, U.K. (email: z.ding@lancaster.ac.uk).}
\thanks{L. Song is with the School of Electrical Engineering and Computer Science, Peking University, Beijing 100871, China (e-mail: lingyang.song@pku.edu.cn).}}

\maketitle

\begin{abstract}
This paper considers the antenna selection (AS) problem for a MIMO non-orthogonal multiple access (NOMA) system. In particular, we develop new computationally-efficient AS algorithms for two commonly-used scenarios: NOMA with fixed power allocation (F-NOMA) and NOMA with cognitive radio-inspired power allocation (CR-NOMA), respectively. For the F-NOMA scenario, a new max-max-max antenna selection (A$^3$-AS) scheme is firstly proposed to maximize the system sum-rate. This is achieved by selecting one antenna at the base station (BS) and corresponding best receive antenna at each user that maximizes the channel gain of the resulting \textit{strong} user. To improve the user fairness, a new max-min-max antenna selection (AIA-AS) scheme is subsequently developed, in which we jointly select one transmit antenna at BS and corresponding best receive antennas at users to maximize the channel gain of the resulting \textit{weak} user. For the CR-NOMA scenario, we propose another new antenna selection algorithm, termed maximum-channel-gain-based antenna selection (MCG-AS), to maximize the achievable rate of the secondary user, under the condition that the primary user's quality-of-service requirement is satisfied. The asymptotic closed-form expressions of the average sum-rate for A$^3$-AS and AIA-AS and that of the average rate of the secondary user for MCG-AS are derived, respectively. Numerical results demonstrate that the AIA-AS provides better user-fairness, while the A$^3$-AS achieves a near-optimal sum-rate in F-NOMA systems. For the CR-NOMA scenario, MCG-AS achieves a near-optimal performance in a wide signal-to-noise-ratio regime. Furthermore, all the proposed AS algorithms yield a significant computational complexity reduction, compared to exhaustive search-based counterparts.
\end{abstract}\vspace{-0.5em}


\IEEEpeerreviewmaketitle

\section{Introduction}
The non-orthogonal multiple access (NOMA) technique has emerged as a promising solution to significantly improve the spectral efficiency of the next-generation wireless networks \cite{Ref_Saito,Ref_NOMA,Ref_NOMA_2}. By superimposing the information of multiple users in the power domain, multiple users can be served within the same time/frequency/code domain. Different from the conventional water-filling power allocation strategy, the NOMA technique generally allocates more transmit power to the users with the poor channel conditions (\textit{weak} users). In this case, these users can decode their higher-power-level signals directly by treating others' signals as noise. In contrast, those users with the better channel conditions (\textit{strong} users) adopt the successive interference cancellation (SIC) technique for signal detection. It has been demonstrated that both the throughput performance and user fairness can be significantly improved in NOMA systems compared to conventional orthogonal multiple access (OMA) systems~\cite{Ref_Saito,Ref_NOMA_improvement}.

Initial efforts on the design and analysis of the NOMA technique focused on single-antenna systems, see e.g., \cite{Ref_Co_NOMA,Ref_uplink_NOMA,Ref_relay,Ref_power_allocation,Ref_Yuanwei_Liu,Ref_Henry1,Ref_Henry2}. Recently, multiple antennas have been used in NOMA systems (MIMO-NOMA) to exploit the spatial degrees of freedom~\cite{Ref_MIMO1,Ref_MIMO2,Ref_Massive_MIMO,Ref_MIMO_3,Ref_MIMO_4}. Although the system capacity can be potentially scaled up with the number of antennas, this superior performance comes at the price of expensive RF chains, high power consumption and computational complexity required for signal processing at both the transmitter and receivers. To avoid the high hardware costs, demanding power consumption and heavy computational burden while preserving the diversity and throughput benefits from MIMO, the antenna selection (AS) technique has been recognized as an effective solution \cite{Ref_AS,Ref_AS2,Ref_AS3}. There are only a few papers that considered the AS problem for MIMO-NOMA systems in the open literature. Specifically, a transmit AS (TAS) algorithm was proposed in \cite{Ref_TAS_in_NOMA}, and a joint TAS and user scheduling algorithm was considered in \cite{Ref_AS_in_Massvie_MIMO_NOMA}. However, both \cite{Ref_TAS_in_NOMA} and \cite{Ref_AS_in_Massvie_MIMO_NOMA} only focused on the TAS design at the base station (BS) side and there were no analytical characterization of the system performance for the proposed algorithms.

To our best knowledge, joint AS at both the BS and users for MIMO-NOMA systems is still an open problem. Though there have been some published results on joint AS schemes in conventional MIMO-OMA systems, they cannot be extended to MIMO-NOMA systems directly. This is mainly because there is severe inter-user interference in MIMO-NOMA systems while the signals are transmitted in an interference-free manner in MIMO-OMA systems. A global optimal solution of this problem requires an exhaustive search (ES) over all possible antenna combinations, and its complexity would become unacceptable when the numbers of antennas at all terminals become large. Motivated by this, in this paper, we focus on the design and analysis of the low-complexity joint AS algorithms for a typical MIMO-NOMA downlink scenario, where a small number of users (e.g., two users) are grouped together to perform NOMA and different groups are admitted into the network in a OMA manner \cite{Ref_Saito,Ref_MIMO2,Ref_power_allocation1}. In particular, two commonly used power allocation policies: NOMA with fixed power allocation (F-NOMA) and NOMA with cognitive radio-inspired power allocation (CR-NOMA)  \cite{Ref_power_allocation1,Ref_power_allocation2,Ref_power_allocation3}, are considered respectively. In F-NOMA, the set of power allocation coefficients is predefined and dynamically allocated to users in each resource block. In contrast, in CR-NOMA, the quality-of-service (QoS) requirements of users are satisfied with different priorities. Generally, the user with lower priority (i.e., secondary user, SU) is opportunistically served only when the QoS of user with higher priority (i.e., primary user, PU) is guaranteed. In this case, the set of power allocation coefficients in CR-NOMA is not predetermined but depending on the QoS requirement of the PU and the instantaneous channel order of users. Note that the design of low-complexity AS algorithm for MIMO-NOMA systems is fundamentally different from that for conventional MIMO-OMA systems. This is because multiple channel matrices with different distributions need to be considered concurrently to perform the antenna selection in NOMA, while only the channel status of one user is required for OMA. Furthermore, selecting different antenna not only affects the instantaneous channel gains of individual users, but also easily leads to different order of instantaneous channel gains of all users. This means that the information decoding order at the user side and the resulting expressions of the signal-to-interference-plus-noise (SINR) of different users are not predetermined but largely affected by the antenna selection process, which makes the design and analysis of antenna selection for MIMO-NOMA system non-trivial. The main contributions of this paper are summarized as~follows:

\begin{itemize}
  \item For the F-NOMA scenario,  a max-max-max AS (A$^3$-AS) algorithm is firstly designed to maximize the system sum-rate. This is achieved by selecting one antenna at the BS and corresponding best receive antenna at each user that maximize the channel gain of the resulting \textit{strong} user. To improve the user fairness, a new max-min-max antenna selection (AIA-AS) scheme is subsequently developed, in which one transmit antenna at BS and corresponding best receive antennas at users are jointly selected to maximize the channel gain of the resulting \textit{weak} user.
  \item For the CR-NOMA scenario, a joint antenna selection scheme called maximum-channel-gain-based AS (MCG-AS) algorithm is first proposed to maximize the achievable rate of the SU subject to the QoS requirements of the PU, and two simplified versions of MCG-AS, i.e., primary-user-based AS (PU-AS) and second-user-base AS (SU-AS), are subsequently presented to further reduce the computational complexity of MCG-AS.
  \item Asymptotic closed-form expressions for the average sum-rates (average rate of the SU) of the proposed algorithms in F-NOMA (CR-NOMA) systems are derived for high SNR scenarios,  respectively.
  \item The computational complexities for all the proposed joint AS algorithms are considered, and are shown to be significantly lower than that of the ES-based schemes. Extensive simulation results illustrate that both A$^3$-AS and AIA-AS in F-NOMA systems yield significant performance gains over the OMA ES-based scheme and the random selection scheme. In particular, A$^3$-AS can achieve the near-optimal system sum-rate while AIA-AS can provide better user fairness. In CR-NOMA, the MCG-AS scheme can achieve the near-optimal rate performance in a wide SNR regime. In contrast, the rate performance of both PU-AS and SU-AS is significantly affected by the distances between the BS and users. Specifically, the PU-AS (SU-AS) scheme can approach the optimal performance when the primary (secondary) user is closer to the BS.
\end{itemize}

The rest of the paper is organized as follows. In Section II, the considered MIMO-NOMA downlink model is introduced, and the joint AS optimization problems for both F-NOMA and CR-NOMA scenarios are formulated, respectively. In Section III, two new joint AS schemes for F-NOMA systems are proposed and their achievable sum-rate are analysed. In Section IV, a computationally-efficient joint AS algorithm and its two simplified versions are developed and analysed for CR-NOMA systems. Finally, simulation results and comparisons between the proposed algorithms and benchmark schemes are given in Section V, and conclusions are drawn in~Section VI.

\emph{Notations}: ${\mathcal{C}^{m \times n}}$ represents the set of all $m\times n$ matrices. All bold uppercase letters represent matrices, and all Calligraphic letters represent sets. $\mathrm{Pr}(\cdot)$ denotes the probability of an event, $\log\left(\cdot\right)$ denotes the logarithm function based on $2$, and $\left|{\cdot}\right|$ and $\mathbb{E}[\cdot]$ denote the absolute value and the expectation operation, respectively. $F_X(x)$ and $f_X(x)$ represent the cumulative density function (CDF) and the probability density function~(PDF) of a random variable $X$, respectively. $\mathcal{O}$ is usually used in the efficiency analysis of algorithms and $q(x)=\mathcal{O}(p(x))$ when $\lim\limits_{x\rightarrow\infty}|\frac{q(x)}{p(x)}|=c, 0<c<\infty$.
\vspace{-1em}
\section{System model}
Consider a downlink communication scenario with one BS and multiple users. As discussed in \cite{Ref_power_allocation1}, it is impractical to ask all the users in a system to perform NOMA due to the strong co-channel interference and heavy system overhead for user coordination. One promising alternative is to construct a hybrid multiple access system, wherein the users are divided into multiple small groups, and NOMA is implemented within each group while OMA is performed between different groups. Without loss of generality, in this paper, we follow \cite{Ref_Saito,Ref_MIMO2,Ref_power_allocation1} and assume two users\footnote{Note that the two-user setting is how NOMA is implemented in LTE-Advanced.} are paired in each group randomly to perform NOMA and different groups are admitted into the network in a time-division multiple access manner. There is no doubt that some other delicately designed user pairing schemes may further improve the rate performance of the NOMA implementation, however, it is out of the scope of~this~paper.

We consider that the BS is equipped with $N$ antennas, while the two selected users (termed as UE$_1$ and UE$_2$) are equipped with $M$ and $K$ antennas, respectively. The channel matrix from the BS to UE$_1$ (UE$_2$) is denoted by ${\bf{H}} \in {\mathcal{C}^{N \times M}}$ (${\bf{G}} \in {\mathcal{C}^{N \times K}}$). We assume that the channels between the BS and users undergo spatially independent flat Rayleigh fading. The entries of $\mathbf{H}$ ($\mathbf{G}$), e.g., ${\tilde{h}_{nm}}$ (${\tilde{g}_{nk}}$), can be modelled as independent and identically distributed (i.i.d.) complex Gaussian random variables, where $\tilde{h}_{nm}$ ($\tilde{g}_{nk}$) represents the channel coefficient between the $n$th antenna of the BS and the $m$th ($k$th) antenna of UE$_1$ (UE$_2$). For notation simplicity, we define ${h}_{nm}=|\tilde{h}_{nm}|^2$ and ${g}_{nk}=|\tilde{g}_{nk}|^2$.

As in \cite{Ref_one_RF1,Ref_one_RF2}, we assume that the BS selects one (e.g., $n$th) out of $N$ available antennas to transmit information, while the users select one (e.g., $m$th and $k$th) out of $M$ and $K$ available antennas to receive messages, respectively. In this case, only one RF chain is needed at each node such that the hardware cost and complexity can be reduced. Moreover, only the partial channel state information (CSI), i.e., the channel amplitudes, is needed instead of the full CSI at the BS, which costs much less feedback bandwidth\footnote{It is worth noting that for the case of slow fading (i.e., the channel coherent time is long such that the time used for channel estimation can be neglected compared to the duration of each transmission block), the obtained results in this paper can serve as a tight performance bound for the case of considering channel estimation overhead. When the channels change relatively quickly, the impact of channel estimation overhead can be significant. However, incorporating the channel estimation cost into the considered MIMO-NOMA system will require the definition of new system performance metric, the joint optimization of channel estimation overhead, antenna selection, and power allocation, which could constitute a new full paper and thus has been left as a future work.}.

Let $\delta_{n,m,k}$ denote the channel order indicator, defined as
\begin{eqnarray}\label{Equ_part1_Delta}
\delta_{n,m,k}=\left\{ {\begin{array}{*{20}{c}}
{1,}&{\mathrm{if}~{h}_{nm}\ge {g}_{nk}},\\
{0,}&{\mathrm{if}~{h}_{nm}< {g}_{nk}}.
\end{array}} \right.
\end{eqnarray}

According to the NOMA principle, the BS broadcasts the signals superimposed in the power domain~as
\begin{eqnarray}\label{Equ_part1_x}
x=\delta_{n,m,k}(\sqrt{b}s_1 + \sqrt{a}s_2)+({1-\delta_{n,m,k} })(\sqrt{a}s_1 + \sqrt{b}s_2),
\end{eqnarray}where $s_i$ denotes the signal to UE$_i$ with $\mathbb{E}[{\left| {{s_i}} \right|^2}]=1$, and $a$ and $b$ are the power allocation coefficients with $a+b=1$. Without loss of generality, we assume that $a>b$ to guarantee that more power is allocated to the instantaneous \textit{weak}~user.

The received signals at users are given by
\begin{eqnarray}\label{Equ_part1_y}
{y_1}=\sqrt{P_s}\tilde{h}_{nm}x+n_1,~~{y_2}=\sqrt{P_s}\tilde{g}_{nk}x+n_2,
\end{eqnarray}where $P_s$ is the transmit power at the BS, and $n_i$ is the complex additive white Gaussian noise with variance $\sigma_i^2$. For notational simplicity, we hereafter assume $\sigma_1^2 = \sigma_2^2=\sigma^2$.

When $\delta_{n,m,k}=1$, UE$_2$ is the \textit{weak} user and UE$_1$ is the \textit{strong} user, hence the power level of $s_2$ is larger than that of $s_1$. In this case, UE$_2$ decodes $s_2$ directly by treating $s_1$ as noise. In contrast, UE$_1$ first decodes $s_2$ and subtracts it via SIC, and then decodes its own $s_1$ without interference. For the case $\delta_{n,m,k}=0$, the decoding order is reversed. By using the fact that the channels are ordered, it can be easily verified that SIC can be implemented successfully, and the following two rates can be achievable to UE$_1$ and UE$_2$, respectively:
\begin{small}
\begin{eqnarray}\label{Equ_part1_R1}
{R_1}\!\!\!\!&=&\!\!\!\!\delta_{n,m,k}\log\left(1+\rho b{h}_{nm}\right)\!+\!\left(1\!-\!\delta_{n,m,k}\right)\log\!\!\left(1\!+\!\frac{a{h}_{nm}}{b{h}_{nm}+\frac{1}{\rho}}\right),~~~\\
\label{Equ_part1_R2}
{R_2}\!\!\!\!&=&\!\!\!\!\delta_{n,m,k}\log\!\!\left(1+\frac{a{g}_{nk}}{b{g}_{nk}+\frac{1}{\rho}}\right)\!+\!\left(1\!-\!\delta_{n,m,k}\right)\log\left(1\!+\!\rho b{g}_{nk}\right),~~~
\end{eqnarray}\end{small}where ${\rho} = {{{P_s}} \mathord{\left/ {\vphantom {{{P_s}} {{\sigma ^2}}}} \right.\kern-\nulldelimiterspace} {{\sigma ^2}}}$ is the transmit SNR.

To evaluate the user fairness, the Jain's fairness index \cite{Ref_Fairness} is adopted in this paper, which can be expressed as
\begin{eqnarray}\label{Equ_part3_FNOMA_Fairness}
\eta=(R_1+R_2)^2/{2(R_1^2+R_2^2)}.
\end{eqnarray}We can observe that the Jain's fairness index is bounded between 0 and 1 with the maximum achieved by equalling user's~rates. Note that the Jain’s fairness index is a performance metric rather than an approach to guarantee the fairness of the considered MIMO-NOMA systems. To improve user fairness, the interested readers are referred to various power allocation schemes for NOMA, e.g., ordered power allocation \cite{Ref_Fairness_1}, max-min power allocation \cite{Ref_Fairness_2}, and $\alpha$-fairness \cite{Ref_Fairness_3}, etc.

In the following subsections we formulate the joint AS optimization problems for two MIMO-NOMA scenarios, i.e., F-NOMA and CR-NOMA, respectively.
\vspace{-1em}
\subsection{Joint AS Optimization Problem for F-NOMA Systems}
In F-NOMA, fixed power allocation coefficients are predefined. In this case, we can formulate the following optimization problem to maximize the achievable system~sum-rate
\begin{eqnarray}\label{Equ_part2_FNOMA_problem}
\mathbf{P1}:~\left\{ n^*,m^*,k^*\right\}  = \mathop {\arg \max }\limits_{n \in {\mathcal{N}},m \in {\mathcal{M}},k \in {\mathcal{K}}} {R_{\mathrm{sum}}}\left({h}_{nm},{g}_{nk}\right),
\end{eqnarray}where $\mathcal{N}=\{1,2,\cdots,N\}$, $\mathcal{M}=\{1,2,\cdots,M\}$, $\mathcal{K}=\{1,2,\cdots,K\}$ and the achievable sum-rate is given by
\begin{eqnarray}\label{Equ_part2_FNOMA_rsum}
R_{\mathrm{sum}}\left({h}_{nm},{g}_{nk}\right)\!\!\!&=&\!\!\!R_1 +R_2\nonumber\\
\!\!\!&=&\!\!\!\log\left(1+\rho b{\gamma}_n^{s}\right)+\log\!\!\left(1+\frac{a{\gamma}_n^{w}}{b{\gamma}_n^{w}+1/\rho}\right),~~~
\end{eqnarray}where ${\gamma}_n^s=\max\left({h}_{nm},{g}_{nk}\right)$ denotes the channel gain of the \textit{strong} user and ${\gamma}_n^w=\min\left({h}_{nm},{g}_{nk}\right)$ denotes the channel gain of the \textit{weak} user. It is worth pointing out that for a given time slot, which user is the \textit{strong} (\textit{weak}) user is not predefined, but depending on the channel fluctuations and antenna selection~result. This uncertainty needs to be carefully considered when designing the antenna selection algorithm.\vspace{-1em}

\subsection{Joint AS Optimization for CR-NOMA Systems}
In the considered CR-NOMA downlink scenario, without loss of generality, we treat UE$_2$ as the primary user and UE$_1$ as the secondary user. Specifically, UE$_1$ is opportunistically served when the QoS of UE$_2$ is achieved. Hereafter we use $R_1^c$ and $R_2^c$ to respectively denote the achievable rate of UE$_1$ and UE$_2$, where the superscript $(\cdot)^{{c}}$ is used to distinguish the parameters in CR-NOMA systems from those in F-NOMA systems. Mathematically, the achievable rate of the secondary user, $R_2^{{c}}$, should satisfy $R_2^{{c}}\geq {R}_{\mathrm{th}}$, where ${R}_{\mathrm{th}}$ is the QoS threshold of UE$_2$. By noting that the power allocation coefficient $0\leq b^c_{n,m,k}\leq 1$, we can express the valid range of $b^c_{n,m,k}$ as below:
\begin{eqnarray}\label{Equ_part2_CR_b}
      b^c_{n,m,k}= \left\{ {\begin{array}{*{20}{c}}
      {\min\left({\frac{\varepsilon}{\rho{g}_{nk}},~1}\right),}&{\mathrm{if}~\delta_{n,m,k}=0},\\
      {\max\left(\frac{\rho g_{nk}-\varepsilon}{\rho g_{nk}(\varepsilon+1)},~0\right),}&{\mathrm{if}~\delta_{n,m,k}=1},
\end{array}} \right.
\end{eqnarray}where $\varepsilon=2^{{R}_{\mathrm{th}}}-1$, and we then have $a^c_{n,m,k}=1-b^c_{n,m,k}$.

By substituting (\ref{Equ_part2_CR_b}) into (\ref{Equ_part1_R1}), we can observe that the achievable rate of the secondary user, $R_1^{{c}}$, is zero when $\delta_{n,m,k}=0,~b^c_{n,m,k}=1$ or $\delta_{n,m,k}=1,~b^c_{n,m,k}=0$. In other words, when the QoS of UE$_2$ cannot be satisfied, no power would be allocated to UE$_1$ and hence $R_1^{{c}}=0$. For the considered high SNR scenarios, i.e., $\rho\rightarrow \infty$, the expression of $b_{n,m,k}^c$ given in (\ref{Equ_part2_CR_b}) can be simplified as
\begin{eqnarray}\label{Equ_part2_CR_b_simple}
      b_{n,m,k}^c= \left\{ {\begin{array}{*{20}{c}}
      {\frac{\varepsilon}{\rho{g}_{nk}},}&{\mathrm{if}~\delta_{n,m,k}=0},\\
      {\frac{\rho g_{nk}-\varepsilon}{\rho g_{nk}(\varepsilon+1)},}&{\mathrm{if}~\delta_{n,m,k}=1}.
\end{array}} \right.
\end{eqnarray}

By substituting (\ref{Equ_part2_CR_b_simple}) into (\ref{Equ_part1_R1}), the achievable rate of the secondary user (i.e., $R_1^{{c}}$) can be represented as
\begin{small}
\begin{eqnarray}\label{Equ_part2_CR_R1}
R_1^{{c}}\left(h_{nm},g_{nk}\right)\!\!\!\!\!&=&\!\!\!\!\!\delta_{n,m,k}\log\left(1-\frac{\varepsilon h_{nm}}{(\varepsilon+1)g_{nk}}+\frac{\rho h_{nm}}{\varepsilon+1}\right)\nonumber\\
\!\!\!\!\!&+&\!\!\!\!\!(1-\delta_{n,m,k})\log\!\left(\frac{g_{nk}}{\varepsilon h_{nm}+g_{nk}}+\frac{\rho h_{nm}g_{nk}}{\varepsilon h_{nm}+g_{nk}}\right)\nonumber\\
\!\!\!\!\!&\overset{\rho\rightarrow \infty}{\approx}&\!\!\!\!\!\delta_{n,m,k}\log\left(\frac{\rho h_{nm}}{\varepsilon+1}\right)\nonumber\\
\!\!\!\!\!&+&\!\!\!\!\!(1-\delta_{n,m,k})\log\!\!\left(\frac{\rho h_{nm}g_{nk}}{\varepsilon h_{nm}+g_{nk}}\right).
\end{eqnarray}\end{small}

Now we can formulate the following joint AS problem for CR-NOMA systems:
\begin{eqnarray}\label{Equ_part2_CR_problem}
      \mathbf{P2}:~\{ n^*,m^*,k^*\} & =&\mathop {\arg \max }\limits_{n \in {\mathcal{N}},m \in {\mathcal{M}},k \in {\mathcal{K}}} {R_{1}^{{c}}}({h}_{nm},{g}_{nk})~~\\
      &&\mathrm{s.t.}~R_2^c\left(h_{nm}, g_{nk}\right)\geq R_{\mathrm{th}}.\nonumber
\end{eqnarray}
It is straightforward to see that both $\mathbf{P1}$ and $\mathbf{P2}$ are NP-hard problems, which mean finding the optimal combinations of antennas at both the BS and users may require an exhaustive search with the complexity of $\mathcal{O}\left( {NMK} \right)$. This becomes unaffordable when $N$, $M$ and $K$ become large. Motivated by this, in the subsequent two sections we will develop joint AS algorithms for F-NOMA and CR-NOMA systems with significantly reduced computational complexity,~respectively.\vspace{-1em}

\section{Antenna Selection for F-NOMA systems}
In this section, two low-complexity joint AS algorithms are developed to maximize the system sum-rate for F-NOMA systems without and with the consideration of user fairness. The closed-form expressions for the average sum-rate of these two proposed algorithms are derived in high SNR regime.

\subsection{Proposed AS Algorithms for F-NOMA Systems}
We can readily verify that the sum-rate $R_{\mathrm{sum}}$ defined in (\ref{Equ_part2_FNOMA_rsum}) is a monotonically increasing function of the channel gains of both the \textit{strong} and \textit{weak} user (i.e.,  ${\gamma}_n^s$ and ${\gamma}_n^w$). Unfortunately, ${\gamma}_n^s$ and ${\gamma}_n^w$ cannot be maximized at the same time in most cases. This is because both the two users need to share the same transmit antenna of the BS and it is very less likely that selecting one of the transmit antennas can lead to the maximum channel gains of both users. Based on this observation, we propose two new joint AS schemes for the considered F-NOMA systems in this subsection, termed max-max-max AS (A$^3$-AS) and max-min-max AS (AIA-AS), respectively. The key idea of A$^3$-AS (AIA-AS) is to make $\gamma_n^w$ ($\gamma_n^s$) as large as possible under the condition that $\gamma_n^s$ ($\gamma_n^w$) is maximized.  More details are elaborated as below.

\subsubsection{A$^3$-AS}
A$^3$-AS mainly consists of three stages.

\begin{itemize}
\item {\textbf{Stage 1}}. For a given transmit antenna, we firstly find the best receive antennas for UE$_1$ and UE$_2$, respectively. Mathematically, we need to find out the largest element ${h_{n}^{\max}}$ (${g_{n}^{\max}}$) in each row of ${\bf{H}}$ (${\bf{G}}$) as follows:
  \begin{eqnarray}\label{Equ_part3_A3AS_Stage11}
  {{h}_{n}^{\max}} &=& \max \left({h}_{n1},\cdots,{h}_{nM}\right),\\
  \label{Equ_part3_A3AS_Stage12}
  {{g}_{n}^{\max}} &=& \max \left({g}_{n1},\cdots,{g}_{nK}\right),
  \end{eqnarray}for $n\in\mathcal{N}$. Each pair $({h}_{n}^{\max},{g}_{n}^{\max})$ is then treated as one AS candidate. The set of all the $N$ pairs can be written as $\mathcal{S}^{(1)}_{A^3}=\left\{({h}_{1}^{\max},{g}_{1}^{\max}),\cdots,({h}_{N}^{\max},{g}_{N}^{\max})\right\}$.

  \item \textbf{Stage 2}. For a given pair from $\mathcal{S}^{(1)}_{A^3}$, the one with larger channel gain is set as the \textit{strong} user while the other one is set as the \textit{weak} user. In order to maximize the channel gain of the \textit{strong} user, we need to find out all its potential channel gains. Mathematically, we need to find out the \textit{larger} element of $(h_{n}^{\max },g_{n}^{\max })$ as follows:
  \begin{eqnarray}\label{Equ_part3_A3AS_Stage_2}
  \gamma_{n}^s= \max ({h_{n}^{\max }},{g_{n}^{\max }}),~~n\in{\mathcal{N}}.
  \end{eqnarray}The set of the $N$ \textit{larger} elements are denoted by $\mathcal{S}^{(2)}_{A^3}=\{\gamma_{1}^s,\cdots,\gamma_{N}^s\}$.

  \item \textbf{Stage 3}. After obtaining the set of the potential channel gains of the \textit{strong} user, we thirdly find out the largest one from $\mathcal{S}^{(2)}_{A^3}$ to maximize $\gamma_n^s$. Mathematically, we have
 \begin{eqnarray}\label{Equ_part3_A3AS_Stage_3}
  \gamma^s_{A^3}= \max (\gamma_{n}^s),~~n\in{\mathcal{N}}.
  \end{eqnarray}
\end{itemize}

Recall that $\gamma^s_{A^3}$ coming from UE$_1$ or UE$_2$ is not predetermined, we then use ($n_{A^3}^*$, $m_{A^3}^*$) to denote the original row and column indexes of $\gamma^s_{A^3}$ when it lies in $\bf H$. In this case, the $n_{A^3}^*$th antenna at the BS and the $m_{A^3}^*$th antenna at UE$_1$ are selected. We use $k_{A^3}^*$ to denote the original column index of $\gamma^w_{A^3}=g_{n_{A^3}^*}^{\max}$. Therefore the $k_{A^3}^*$th antenna at UE$_2$ can be selected concurrently. For the case that $\gamma^s_{A^3}$ lies in $\bf G$, the selected antenna indexes can be obtained in a similar~way.

In A$^3$-AS, the channel gain of the instantaneous \textit{strong} user $\gamma^s_{A^3}$ is maximized. However, the achievable rate of the instantaneous \textit{weak} user cannot be guaranteed and may be worse for some cases. Recall that the Jain's fairness index decreases when the gap between the achievable rates of users enlarges. In order to improve the user fairness, we subsequently develop a AIA-AS scheme, which aims to improve the rate performance of the instantaneous \textit{weak} user and to narrow the gap of the rates of users.
\vspace{-1em}
\subsubsection{AIA-AS}
Similar to the A$^3$-AS algorithm, the AIA-AS scheme also has three stages. The main difference between A$^3$-AS and AIA-AS lies in the second and the third stages. In particular, A$^3$-AS constructs the set of all potential channel gains for the \textit{strong} user and then maximizes $\gamma_n^s$. In contrast, AIA-AS tries to obtain the set of all potential channel gains for the \textit{weak} user and then to maximize $\gamma_n^w$. The three stages of AIA-AS are elaborated as follow.
\begin{itemize}
\item \textbf{Stage 1}. Construct the set $\mathcal{S}^{(1)}_{AIA}=\{(h_{1}^{\max },g_{1}^{\max }),\cdots,(h_{N}^{\max },g_{N}^{\max })\}$ as (\ref{Equ_part3_A3AS_Stage11}) and (\ref{Equ_part3_A3AS_Stage12}).
\item {\textbf{Stage 2}}. In order to maximize the channel gain of the \textit{weak} user, we need to find out all the potential channel gains of the \textit{weak} user from $\mathcal{S}^{(1)}_{AIA}$. Mathematically, we need to find out the \textit{smaller} element in each pair of $(h_{n}^{\max },g_{n}^{\max })$, that is,~
 \begin{eqnarray}\label{Equ_part3_AIA_Stage2}
  \gamma_n^w= \min ({h}_{n}^{\max},{g}_{n}^{\max}),~~n\in{\mathcal{N}}.
  \end{eqnarray}The set containing these $N$ smaller elements is denoted by $\mathcal{S}^{(2)}_{AIA}=\{\gamma_1^w,\cdots,\gamma_N^w\}$.
\item {\textbf{Stage 3}}. After obtaining the set of the potential channel gains of the \textit{weak} user, we thirdly find out the largest one from $\mathcal{S}^{(2)}_{AIA}$ to maximize $\gamma_n^w$. Mathematically, we have
  \begin{eqnarray}\label{Equ_part3_AIA_Stage3}
  \gamma^w_{AIA}= \max (\gamma_n^w),~~n\in{\mathcal{N}}.
  \end{eqnarray}
\end{itemize}

Similarly, we can see that $\gamma^w_{AIA}$ coming from UE$_1$ or UE$_2$ is not predefined. We use ($n_{AIA}^*$, $k_{AIA}^*$) to denote the original row and column indexes of $\gamma^w_{AIA}$ when it lies in $\bf G$. In this case, the $n_{AIA}^*$th antenna at the BS and the $k_{AIA}^*$th antenna at UE$_2$ are selected, respectively. Meanwhile, we use $m_{AIA}^*$ to denote the original column index of $\gamma^s_{AIA}={h}_{n_{AIA}^*}^{\max}$. Therefore the $m_{AIA}^*$th antenna at UE$_1$ can be selected simultaneously. For the case that $\gamma^w_{AIA}$ lies in $\bf H$, the selected antenna indexes can be obtained in a similar~way.

\subsubsection{Computational Complexities Comparisons}
As mentioned before, the complexity of the optimal selection algorithm achieved by the ES-based scheme in F-NOMA systems is as high as $\mathcal{O}\left(NMK\right)$. When the number of antennas at each node becomes large, e.g., up to hundreds of antennas \cite{Ref_number_of_antennas}, the computational burden would increase significantly.

The two new AS algorithms for F-NOMA systems dramatically reduce the selection complexity to $\mathcal{O}\big(N(M+K+3)\big)$, where the main computation lies in sorting the channel gains. For the case $N=M=K$, we can find that the complexities of A$^3$-AS and AIA-AS are approximately $\mathcal{O}(N^2)$, which is an order of magnitude lower compared to that of the optimal ES scheme, which is $\mathcal{O}(N^3)$.
\vspace{-1em}
\subsection{Performance Analysis of the New Algorithms for F-NOMA}
In this subsection, we analyse the average sum-rate of the proposed joint AS algorithms, i.e., A$^3$-AS and AIA-AS in F-NOMA systems, respectively.

Assuming flat Rayleigh fading, ${h}_{nm}$ is then an exponentially distributed random variable. When $x\geq 0$, the CDF and PDF of $h_{nm}$ are respectively given by
\begin{eqnarray}\label{Equ_part5_h_im_cdf_pdf}
F_{{h}}(x)=1-e^{-\Omega_h x},~~f_{{h}}(x)={\Omega_h}{e^{-\Omega_h x}},
\end{eqnarray}where $\Omega_h=1/\mathbb{E}[h_{nm}]$. Similarly, when $x\geq 0$ we have the CDF and PDF of ${g}_{nk}$ as follows,
\begin{eqnarray}\label{Equ_part5_g_ik_cdf_pdf}
F_{{g}}(x)=1-e^{-\Omega_gx},~~f_{{g}}(x)=\Omega_g{e^{-\Omega_gx}},
\end{eqnarray}where $\Omega_h=1/\mathbb{E}[g_{nk}]$.

Recall that both the two proposed AS schemes for F-NOMA need to find the maximum element in each row of $\bf H$ and $\bf G$ (i.e.,
$h_{n}^{\max}$ and $g_{n}^{\max}$). We thus first obtain the distribution functions of $h_{n}^{\max}$ for $x\geq 0$ as below,
\begin{eqnarray}\label{Equ_part5_himax_cdf}
F_{h_{n}^{\max}}(x)\!\!\!&=&\!\!\!\left(1-e^{-\Omega_h x}\right)^{M}\!\overset{(c_1)}{=}\!\sum\nolimits_{i=0}^M\mu_{i,M}e^{-i\Omega_hx},\\
\label{Equ_part5_himax_pdf}
f_{h_{n}^{\max}}(x)\!\!\!&=&\!\!\!-\sum\nolimits_{i=1}^Mi\Omega_h\mu_{i,M}e^{-i\Omega_hx},
\end{eqnarray}where $\mu_{i,M}=(-1)^i\binom{M}{i}$ and the expansion step $(c_1)$ is conducted based on the Binomial theorem.

Similarly, the CDF and PDF of $g_{n}^{\max}$ for $x\geq0$ are given~by
\begin{eqnarray}\label{Equ_part5_gimax_cdf}
F_{g_{n}^{\max}}(x)\!\!\!&=&\!\!\!\left(1-e^{-\Omega_g x}\right)^{K}=\sum\nolimits_{j=0}^K\mu_{j,K}e^{-j\Omega_gx},\\
\label{Equ_part5_gimax_pdf}
f_{g_{n}^{\max}}(x)\!\!\!&=&\!\!\!-\sum\nolimits_{j=1}^Kj\Omega_g\mu_{j,K}e^{-j\Omega_gx}.
\end{eqnarray}
From the sum-rate expression of the considered F-NOMA system in (\ref{Equ_part2_FNOMA_rsum}), we can observe that the achievable rate of the instantaneous \textit{weak} UE approaches a constant when $\rho\rightarrow \infty$, i.e., $\log\left(1+\frac{a{\gamma}_n^{w}}{b{\gamma}_n^{w}+1/\rho}\right)\overset{\rho\rightarrow \infty}{\approx}\log\left(1+\frac{a}{b}\right)={\log\left(1/b\right)}$ with $a+b=1$, which is only related to the predefined power allocation coefficient $b$. In this case, we mainly focus on the contribution of the instantaneous \textit{strong} user to the sum-rate. Mathematically, we have the approximation of the system instantaneous sum-rate given in (\ref{Equ_part2_FNOMA_rsum}) as follows:
\begin{eqnarray}\label{Equ_part5_FNOMA_rsum_app}
{R}_{\mathrm{sum}}\left({\gamma}^s_n\right)&\overset{\rho\rightarrow\infty}{\approx}&\log(1+b\rho{\gamma}_n^s)+\log\left(1/b\right).
\end{eqnarray}Subsequently, we perform the asymptotic analysis for the average sum-rates of A$^3$-AS and AIA-AS algorithms, respectively.

\subsubsection{Average Sum-Rate Analysis for A$^3$-AS}
Let $h^{\max}=\max\left(h_{nk}\right)$ and $g^{\max}=\max\left(g_{nk}\right)$ for $n\in\mathcal{N}$, $m\in\mathcal{M}$ and $k\in\mathcal{K}$. Recall that in A$^3$-AS, $\gamma^s_{A^3}$ is actually the relatively larger element of $h^{\max}$ and $g^{\max}$. In order to obtain the distribution of $\gamma^s_{A^3}$, at first we derive the CDF and PDF of $h^{\max}$ and $g^{\max}$ respectively as follows:
\begin{eqnarray}\label{Equ_part5_A3_hmax}
F_{h^{\max}}(x)&=&\sum\nolimits_{i=0}^{NM}\mu_{i,NM}e^{-i\Omega_hx},\\
f_{h^{\max}}(x)&=&-\sum\nolimits_{i=1}^{NM}i\Omega_h\mu_{i,NM}e^{-i\Omega_hx},\\
\label{Equ_part5_A3_gmax}
F_{g^{\max}}(x)&=&\sum\nolimits_{j=0}^{NK}\mu_{j,NK}e^{-j\Omega_gx},\\
f_{g^{\max}}(x)&=&-\sum\nolimits_{j=1}^{NK}j\Omega_g\mu_{i,NK}e^{-j\Omega_gx}.
\end{eqnarray}Then the CDF and PDF of $\gamma^s_{A^3}=\max(h^{\max},g^{\max})$ in A$^3$-AS can be calculated as follows
\begin{small}
\begin{eqnarray}\label{Equ_part5_A3_gamma_s}
F_{\gamma^s_{A^3}}(x)\!\!\!&=&\!\!\!\mathrm{Pr}\left(\max(h^{\max},g^{\max})<x\right)\nonumber\\
\!\!\!&=&\!\!\!\mathrm{Pr}\left(h^{\max}\leq g^{\max}<x)+\mathrm{Pr}(g^{\max}<h^{\max}<x\right),\\
f_{\gamma^s_{A^3}}(x)\!\!\!&=&\!\!\!f_{g^{\max}}(x)\int_0^xf_{h^{\max}}(y)\mathrm{d}y+f_{h^{\max}}(x)\int_0^xf_{g^{\max}}(y)\mathrm{d}y\nonumber\\
\!\!\!&=&\!\!\!\sum_{i=1}^{NM}\sum_{j=1}^{NK}\mu_{i,NM}\mu_{j,NK}\big(i\Omega_he^{-i\Omega_hx}\big.\nonumber\\
\!\!\!&+&\!\!\!\big.j\Omega_ge^{-j\Omega_gx}-(i\Omega_h+j\Omega_g)e^{-(i\Omega_h+j\Omega_g)x}\big).
\end{eqnarray}\end{small}We now can analyse the average sum-rate of A$^3$-AS for high SNR scenarios and obtain Proposition 1 as~below.
\begin{proposition}\label{proposition2_A3_rsum}
With the PDF of $\gamma^s_{A^3}$ in A$^3$-AS, i.e., $f_{\gamma^s_{A^3}}(x)$, when $\rho\rightarrow\infty$, the asymptotic closed-form expression for the average sum-rate of the A$^3$-AS algorithm is given by
\begin{small}
\begin{eqnarray}\label{Equ_lemma3_A3_rsum}
\bar{R}_{\mathrm{sum}}^{A^3}\!\!\!&\approx&\!\!\!\frac{1}{\ln2}\left(\ln\rho-C\right)\nonumber\\
\!\!\!&+&\!\!\!\sum_{i=1}^{NM}\sum_{j=1}^{NK}\frac{(-1)^{i+j}\binom{NM}{i}\binom{NK}{j}}{\ln2}\ln\left(\frac{i\Omega_h+j\Omega_g}{ij\Omega_h\Omega_g}\right)~~~
\end{eqnarray}\end{small}
\end{proposition}
\begin{IEEEproof}
See Appendix A.
\end{IEEEproof}As can be observed from (\ref{Equ_lemma3_A3_rsum}), the system sum-rate $\bar{R}_{\mathrm{sum}}^{A^3}$ is an increasing function of the transmit SNR $\rho$ as expected. Interestingly, it is not affected by the power allocation coefficient $b$. The reason behind this observation is that when $\rho\rightarrow\infty$, ${R}_{\mathrm{sum}}\approx\log(\gamma^s\rho)$, in which the power allocation factor $b$ has been removed. It will be verified via the computer simulations in section V.

\subsubsection{Average Sum-rate Analysis for AIA-AS}
In this subsection, we derive the asymptotic average sum-rate of the proposed AIA-AS algorithm, i.e., $\bar{R}_{\mathrm{sum}}^{AIA}$, for high SNR scenarios. According to (\ref{Equ_part5_FNOMA_rsum_app}), we need to know the PDF of the channel gain of the instantaneous \textit{strong} user who shares the transmit antenna with the \textit{weak} user, i.e., $f_{\gamma^s_{AIA}}(x)$. With the help of the law of full probability and the multinomial theorem, the PDF of $f_{\gamma^s_{AIA}}(x)$ is provided in Lemma 1 as below.
\begin{lemma}\label{Lemma1_AIA_gamma_dis}
In the AIA-AS algorithm, the PDF of the channel gain of the selected \textit{strong} user, i.e., $\gamma^s_{AIA}$, is given by
\begin{small}
\begin{eqnarray}\label{Equ_part5_AIA_gamma_s_pdf}
f_{\gamma^s_{AIA}}(x)=\sum_{i=1}^M\sum_{j=1}^K\sum_{\ell}C_\ell t_\ell \zeta_{ij}\big(\psi(i\Omega_h,j\Omega_g)+\psi(j\Omega_g,i\Omega_h)\big),
\end{eqnarray}\end{small}where
\begin{small}
\begin{eqnarray}
C_\ell&=&\frac{(N-1)!}{\ell_0!\cdots\ell_{MK}!}~~\mathrm{for}~~\ell_0+\cdots+\ell_{MK}=N-1, \nonumber\\
t_{\ell}&=&\prod_{\substack{1\leq i\leq M\\1\leq j\leq K}} (-\mu_{i,M}\mu_{j,K})^{\ell_{ij}},\nonumber\\
\zeta_{ij}&=&Nij\Omega_h\Omega_g\mu_{i,M}\mu_{j,K},\nonumber\\
\psi(\theta_1,\theta_2)&=&e^{-\theta_1x}\left(\frac{e^{-\theta_2 x}-1}{\theta_2}-\frac{e^{-(\theta_2+\xi_\ell) x}-1}{\theta_2+\xi_\ell}\right),\nonumber\\
\xi_{\ell}&=&\sum_{i=1}^M\sum_{j=1}^K(i\Omega_h+j\Omega_j)\ell_{ij}.\nonumber
\end{eqnarray}\end{small}
\end{lemma}
\begin{IEEEproof}
See Appendix B.
\end{IEEEproof}Provided the PDF of $\gamma_{AIA}^s$ in AIA-AS algorithm, we can approximate the average sum-rate achieved by AIA-AS in Proposition \ref{proposition1_BW_rsum} as below.
\begin{proposition}\label{proposition1_BW_rsum}
With the PDF of $\gamma^s_{AIA}$, when $\rho\rightarrow \infty$, the asymptotic average sum-rate of AIA-AS, i.e., $\bar{R}_{\mathrm{sum}}^{AIA}$, is given~by
\begin{small}
\begin{eqnarray}\label{Equ_part5_BW_rsum}
\bar{R}_{\mathrm{sum}}^{AIA}\!\!\!&\approx&\!\!\!\log\frac{1}{b}+\sum_{i=1}^M\sum_{j=1}^K\sum_{\ell}\frac{C_\ell t_\ell}{\ln2}\left(T_1+T_2+T_3+T_4\right),~~~
\end{eqnarray}\end{small}where $T_n,~n=\{1,2,3,4\}$ are given by
\begin{small}
\begin{eqnarray}
T_1&=&\frac{\xi_\ell\tilde{\zeta}_{ij}}{\phi_i}\chi(j\Omega_g),~~~T_2=\frac{\xi_\ell\tilde{\zeta}_{ij}}{\phi_j}\chi(i\Omega_h),\nonumber\\
T_3&=&\frac{{\zeta}_{ij}\phi_{ij,2}\chi(\phi_{ij,1})}{\phi_i\phi_j\phi_{ij,1}},~~~T_4=-\tilde{\zeta}_{ij}\chi(i\Omega_h+j\Omega_g),\nonumber
\end{eqnarray}\end{small}in which, $\tilde{\zeta}_{ij}=N\mu_{i,M}\mu_{j,K}$, $\phi_i=i\Omega_h+\xi_\ell$, $\phi_j=j\Omega_g+\xi_\ell$, $\phi_{ij,1}=i\Omega_h+j\Omega_g+\xi_\ell$, $\phi_{ij,2}=i\Omega_h+j\Omega_g+2\xi_\ell$, $\chi(x)=C+\ln\frac{x}{b\rho}$, and the constant $C$ is the Euler's constant.
\end{proposition}
\begin{IEEEproof}
See Appendix C.
\end{IEEEproof}\vspace{-1em}

\section{Antenna Selection for CR-NOMA systems}
In this section, CR-NOMA is considered, and a computationally efficient joint AS algorithm, i.e., MCG-AS, is proposed to maximize the achievable rate of secondary user UE$_1$. To further reduce the computational complexity, two simplified versions of MCG-AS, i.e., PU-AS and SU-AS, are proposed. In the high SNR regime, the asymptotic expressions of average achievable rates of UE$_1$ using MCG-AS, PU-AS and SU-AS are then derived, respectively.
\vspace{-1em}
\subsection{Proposed AS Algorithms for CR-NOMA Systems}
By observing (\ref{Equ_part2_CR_b})-(\ref{Equ_part2_CR_R1}), we can see that the channel gain of the primary user UE$_2$ (i.e., ${g}_{nk}$) and that of the secondary user UE$_1$ (i.e., ${h}_{nm}$) both make contributions to $R_1^{\mathrm{c}}$ but in different ways in CR-NOMA systems. In particular, if ${g}_{nk}$ increases, less power would be needed to satisfy the QoS requirement of UE$_2$. In other words, more power can be allocated to UE$_1$ and hence $R_1^{c}$ increases. In contrast, the contribution of ${h}_{nm}$ to $R_1^{c}$ does not lie in the power domain, but can boost $R_1^{c}$ directly via providing a better channel gain. However, it is not easy to tell which one, ${g}_{nk}$ or ${h}_{nm}$, has greater contribution to~$R_1^{\mathrm{c}}$. In this case, we propose to maximize the contribution of the larger one from ${g}_{nk}$ and ${h}_{nm}$. Accordingly, a low-complexity AS algorithm, referred to as MCG-AS, is developed for the considered CR-NOMA system. More specifically, the MCG-AS algorithm consists of four stages as~below.
\begin{itemize}
   \item {\textbf{Stage 1}}. Find out the best transmit and receive candidates for UE$_1$ (UE$_2$) which can maximize $h_{nm}$ ($g_{nk}$). Mathematically, we have
  \begin{eqnarray}\label{Equ_part4_CRA_Stage11}
   {{h}^{\max}} &=& \max \left({h}_{n1},\cdots,{h}_{NM}\right),\\
   \label{Equ_part4_CRA_Stage12}
   {{g}^{\max}} &=& \max \left({g}_{n1},\cdots,{g}_{NK}\right).
  \end{eqnarray}
  \item {\textbf{Stage 2}}. Compare and find out the larger one from $h^{\max}$ and $g^{\max}$ and set it as the channel gain of the \textit{strong} user. That is,
  \begin{equation}\label{Equ_part4_CRA_Stage2}
  \gamma^s_{\mathrm{MCG}}=\max(h^{\max},g^{\max}).
  \end{equation}Denote the row index of $\gamma^s_{\mathrm{MCG}}$ as $n_{\mathrm{MCG}}^*$. Then, the $n_{\mathrm{MCG}}^*$th antenna at the BS can be selected.
  \item {\textbf{Stage 3}}. Since the \textit{weak} user share the same transmit antenna with the \textit{strong} user, we should find out the largest element $\gamma^w_{\mathrm{MCG}}$ for the \textit{weak} user from the same $n_{\mathrm{MCG}}^*$th row. Mathematically, we have
   \begin{equation}\label{Equ_part4_CRA_Stage3}
   \gamma^w_{\mathrm{MCG}}\!\!=\!\!\left\{ {\begin{array}{*{20}{c}}
    {\!\!\!\max(g_{n_{\mathrm{MCG}}^*1},\cdots,g_{n_{\mathrm{MCG}}^*K}),}&\!\!\!{\mathrm{if}~\gamma^s_{\mathrm{MCG}}=h^{\max}},\\
    {\!\!\!\max(h_{n_{\mathrm{MCG}}^*1},\cdots,h_{n_{\mathrm{MCG}}^*M}),}&\!\!\!{\mathrm{if}~\gamma^s_{\mathrm{MCG}}=g^{\max}}.
   \end{array}} \right.
   \end{equation}For the case that $\gamma^s_{\mathrm{MCG}}=h^{\max}$, we denote the original column indexes of $\gamma^s_{\mathrm{MCG}}$ and $\gamma^w_{\mathrm{MCG}}$ as $m_{\mathrm{MCG}}^*$ and $k_{\mathrm{MCG}}^*$, respectively. Then, the $m_{\mathrm{MCG}}^*$th antenna at UE$_1$ and the $k_{\mathrm{MCG}}^*$th antenna at UE$_2$ can be selected concurrently. For the case that $\gamma^s_{\mathrm{MCG}}=g^{\max}$, the corresponding antenna indexes at users can be obtained similarly.
  \item {\textbf{Stage 4}}. Compute $b_{n^*,m^*,k^*}^c$ and $a_{n^*,m^*,k^*}^c$ by substituting $\gamma^s_{\mathrm{MCG}}$ and $\gamma^w_{\mathrm{MCG}}$ to (\ref{Equ_part2_CR_b_simple}).
\end{itemize}

We also realize that, when the primary user UE$_2$ is much closer to the BS than the secondary user UE$_1$, it is likely that $g^{\max}$ is larger than $h^{\max}$. In this case, MCG-AS can be simplified into the PU-AS scheme, in which the AS can be performed by maximizing the channel gain of the primary user UE$_2$. In this sense, more power can be allocated to the secondary user UE$_1$. In contrast, when UE$_2$ is much farther from the BS than UE$_1$, it is likely that $g^{\max}$ is smaller than $h^{\max}$. In this case, MCG-AS can be reduced to the SU-AS scheme, which aims to maximize the channel gain of the secondary user UE$_1$ directly. We elaborate the principles and key stages of the simplified PU-AS and SU-AS in the subsequent two subsections.

\subsubsection{PU-AS}
The key idea of PU-AS is to select the antenna pair from the BS and the primary user UE$_2$ concurrently, which provides the maximum channel gain for UE$_2$. In this case, less power is needed for UE$_2$ to satisfy its QoS requirement and more power can thus be allocated to the secondary user UE$_1$ to increase the corresponding $R_1^{c}$. The main three stages of PU-AS scheme are described as below.
\begin{itemize}
   \item {\textbf{Stage 1}}. Find out the best transmit and receive antennas for UE$_2$ to maximize its channel gain. Mathematically, that is to find out the largest element ${g^{\max}}$ from ${\bf{G}}$,
  \begin{eqnarray}\label{Equ_part4_CRP_Stage1}
   {{g}^{\max}} = \max \left({g}_{i1},\cdots,{g}_{NK}\right).
  \end{eqnarray}Denote the row and column indexes of $g^{\max}$ as $n_{\mathrm{P}}^*$ and $k_{\mathrm{P}}^*$, respectively. Then, the $n_{\mathrm{P}}^*$th antenna at the BS and the $k_{\mathrm{P}}^*$th antenna at UE$_2$ are selected simultaneously.
  \item {\textbf{Stage 2}}. For the $n_{\mathrm{P}}^*$th selected transmit antenna, find out the best receive antenna for UE$_1$ who shares the same transmit antenna with UE$_2$. Mathematically, find out the largest element $h_{n_{\mathrm{P}}^*}^{\max}$ in the $n_{\mathrm{P}}^*$th row of $\bf{H}$. That is,
  \begin{eqnarray}\label{Equ_part4_CRP_Stage2}
  h_{n_{\mathrm{P}}^*}^{\max}= \max (h_{n_{\mathrm{P}}^*1},\cdots,h_{n_{\mathrm{P}}^*M}).
  \end{eqnarray}Denote the column index of $h_{n_{\mathrm{P}}^*}^{\max}$ as $m_{\mathrm{P}}^*$. Then, the $m_{\mathrm{P}}^*$th antenna at UE$_1$ is selected.
  \item {\textbf{Stage 3}}. Calculate $b_{n^*,m^*,k^*}^c$ and $a_{n^*,m^*,k^*}^c$ by substituting the value of $g^{\max}$ and $h_{n_{\mathrm{P}}^*}^{\max}$ to (\ref{Equ_part2_CR_b_simple}).
\end{itemize}

\subsubsection{SU-AS}
The key idea of SU-AS is to first select the antenna pair from the BS and secondary user UE$_1$ concurrently, which provides the maximum channel gain for UE$_1$. Similarly, SU-AS consists of three stages.
\begin{itemize}
   \item {\textbf{Stage 1}}. Find out the largest element ${h^{\max}}$ from ${\bf{H}}$.
  \begin{eqnarray}\label{Equ_part4_CRS_Stage1}
   {{h}^{\max}} = \max \left({h}_{i1},\cdots,{h}_{NM}\right).
  \end{eqnarray}Denote the row and column indexes of $h^{\max}$ as $n_{\mathrm{S}}^*$ and $m_{\mathrm{S}}^*$, respectively. Then, the $n_{\mathrm{S}}^*$th antenna at the BS and the $m_{\mathrm{S}}^*$th antenna at UE$_1$ are selected simultaneously.
  \item {\textbf{Stage 2}}. Find out the largest element $g_{n_{\mathrm{S}}^*}^{\max}$ in the $n_{\mathrm{S}}^*$th row of $\bf{G}$. Mathematically,
  \begin{eqnarray}\label{Equ_part4_CRS_Stage2}
  g_{n_{\mathrm{S}}^*}^{\max}= \max (g_{n_{\mathrm{S}}^*1},\cdots,g_{n_{\mathrm{S}}^*K}).
  \end{eqnarray}Denote the column index of $g_{n_{\mathrm{S}}^*}^{\max}$ as $k_{\mathrm{S}}^*$. Then, the $k_{\mathrm{S}}^*$th antenna at UE$_2$ is selected.
  \item {\textbf{Stage 3}}. Obtain $b_{n^*,m^*,k^*}^c$ and $a_{n^*,m^*,k^*}^c$ by substituting the value of $h^{\max}$ and $g_{n_{\mathrm{S}}^*}^{\max}$ to (\ref{Equ_part2_CR_b_simple}).
\end{itemize}

\subsubsection{Computational Complexities Comparisons}
Similar to F-NOMA systems, the complexity of the optimal selection algorithm achieved by the ES-based scheme in CR-NOMA systems is also as high as $\mathcal{O}\left(NMK\right)$. In contrast, all the three proposed AS algorithms, i.e., MCG-AS, PU-AS and SU-AS, can dramatically reduce the selection complexities as illustrated in Table \ref{Tab:Complexity Compair}. Furthermore, the complexities of PU-AS and SU-AS are lower than that of MCG-AS. For the case $N\!=\!M\!=\!K$, the computational burdens of PU-AS and SU-AS are nearly half of that of MCG-AS.
\begin{table}[]
\centering
\caption{Complexity comparison in CR-NOMA}
\label{Tab:Complexity Compair}
\begin{tabular}{@{}ll@{}}
\toprule
           & Computational Complexity \\ \midrule
CR-NOMA-ES & $\mathcal{O}\big(NMK\big)$                        \\
MCG-AS     & $\mathcal{O}\big(N(M+K)+2\big)$                        \\
PU-AS      & $\mathcal{O}\big(NK+M\big)$                        \\
SU-AS      & $\mathcal{O}\big(NM+K\big)$                       \\ \bottomrule
\end{tabular}\vspace{-1em}
\end{table}
\vspace{-1em}
\subsection{Average Rate Analysis of the SU in CR-NOMA}
As discussed above, since the QoS of the primary user UE$_2$ is guaranteed in CR-NOMA systems, we thus only focus on the analysis of the average rate of the secondary user UE$_1$, which is served opportunistically in CR-NOMA networks.

For the proposed MCG-AS scheme, by carefully analysing its antenna selection procedure, we can find that the MCG-AS scheme would become SU-AS when $h^{\max}\geq g^{\max}$ and would simplify to PU-AS otherwise. In this case, the average rate of UE$_1$ in MCG-AS (i.e., $\bar{R}_{1}^{\mathrm{MCG}}$) can be calculated according to the law of total probability as below:
\begin{small}
\begin{eqnarray}\label{Equ_CRA_rsum}
\bar{R}_{1}^{\mathrm{MCG}}\!\!\!&=&\!\!\!\mathrm{Pr}\left(h^{\max}<g^{\max}\right)\bar{R}_1^\mathrm{P}\!+\!\mathrm{Pr}\left(h^{\max}\geq g^{\max}\right)\bar{R}_1^\mathrm{S},
\end{eqnarray}\end{small}where $\bar{R}_1^{\textit{P}}$ and $\bar{R}_1^{\textit{S}}$ denote the average rate of UE$_1$ in PU-AS and SU-AS, respectively, and the probability of the event that $h^{\max}\geq g^{\max}$ is given by
\begin{small}
\begin{eqnarray}\label{Equ_CRA_probability}
\mathrm{Pr}\left(h^{\max}\geq g^{\max}\right)\!\!\!&=&\!\!\!\int_0^\infty f_{h^{\max}}(x)\int_0^x  f_{g^{\max}}(y)\mathrm{d}y\mathrm{d}x\nonumber\\
\!\!\!&=&\!\!\!\sum_{i=1}^{NM}\sum_{j=1}^{NK}{\frac{(-1)^{i+j}\binom{NM}{i}\binom{NK}{j}j\Omega_g}{i\Omega_h+j\Omega_g}}.~~~
\end{eqnarray}\end{small}Accordingly, $\mathrm{Pr}\left(h^{\max}<g^{\max}\right)=1-\mathrm{Pr}\left(h^{\max}\geq g^{\max}\right)$. Therefore, the analytical expression of $\bar{R}_{1}^{\mathrm{MCG}}$ can be obtained if we can respectively derive the expressions of $\bar{R}_1^\mathrm{P}$ and $\bar{R}_1^\mathrm{S}$.

We first calculate $\bar{R}_1^\mathrm{P}$. Recall that in the PU-AS scheme, the maximum element of $\bf G$, i.e., $g^{\max}$, is first selected, and then the maximum element in the corresponding row of $\bf H$ is subsequently selected. By using the i.i.d. property of the all the elements in $\bf H$, the PDF of $h_{n^*_{\mathrm{P}}}^{\max}$ can be derived by using $f_{h_{n}^{\max}}(x)$. In this case, by given the PDF of $g^{\max}$  in (\ref{Equ_part5_A3_gmax}) and that of $h_{n^*_{\mathrm{P}}}^{\max}$ in (\ref{Equ_part5_himax_pdf}), an approximated expression of $\bar{R}_1^\mathrm{P}$ can be can be given in Proposition 3 as below.
\begin{proposition}\label{proposition3_CRP_rsum}
With the QoS requirement of the primary user $R_{\mathrm{th}}$ and the distributions of $g^{\max}$ and $h_{n^*_{\mathrm{P}}}^{\max}$, the average rate of the secondary user UE$_1$ in PU-AS, i.e., $\bar{R}_{1}^{\mathrm{P}}$, can be approximated in the high SNR regime as below:
\begin{small}
\begin{eqnarray}\label{Equ_lemma4_rsum_1}
\bar{R}_{1}^{\mathrm{P}}&\approx&\sum_{i=1}^M\sum_{j=1}^{NK}\frac{(-1)^{i+j}\binom{M}{i}\binom{NK}{j}}{\ln 2}\nonumber\\
&&\left(\frac{\varepsilon j\Omega_g}{i\Omega_h-\varepsilon j\Omega_g}\ln\frac{(\varepsilon+1)j\Omega_g}{i\Omega_h+j\Omega_g}-\ln\frac{i\Omega_h}{\rho}-C\right).~~~
\end{eqnarray}\end{small}
\end{proposition}
\begin{IEEEproof}
See Appendix D.
\end{IEEEproof}

After obtaining $\bar{R}_1^\mathrm{P}$, we thus turn to calculate $\bar{R}_1^\mathrm{S}$ for SU-AS. Recall that in SU-AS, the maximum element of $\bf H$ (i.e., $h^{\max}$) is selected first, followed by the corresponding maximum element in the $n_\mathrm{S}^*$ row of $\bf G$ (i.e., $g_{n^*_\mathrm{S}}^{\max}$). Similarly, considering the i.i.d. property of the all the elements in $\bf G$, the PDF of $g_{n^*_\mathrm{S}}^{\max}$ can be characterized by $f_{g_{n}^{\max}}(x)$. Given the PDF of $h^{\max}$ in (\ref{Equ_part5_A3_hmax}) and that of $g_{n^*_\mathrm{S}}^{\max}$ in (\ref{Equ_part5_gimax_pdf}), the average rate of UE$_1$ in SU-AS, i.e., $\bar{R}_1^\mathrm{S}$, is given by Proposition 4 as below.
\begin{proposition}\label{Proposition4_CRS_rsum}
 With the QoS requirement of the primary UE and the distributions of $h^{\max}$ and $g_{n^*_{\mathrm{S}}}^{\max}$, respectively, the average rate of UE$_1$ in SU-AS, i.e., $\bar{R}_{1}^{\mathrm{S}}$, can be approximated in the high SNR regime as~below:
 \begin{small}
\begin{eqnarray}\label{Equ_lemma5_rsum_1}
\bar{R}_{1}^{\mathrm{S}}&\approx&\sum_{i=1}^{NM}\sum_{j=1}^{K}\frac{(-1)^{i+j}\binom{NM}{i}\binom{K}{j}}{\ln 2}\nonumber\\
&&\left(\frac{\varepsilon j\Omega_g}{i\Omega_h-\varepsilon j\Omega_g}\ln\frac{(\varepsilon+1)j\Omega_g}{i\Omega_h+j\Omega_g}-\ln\frac{i\Omega_h}{\rho}-C\right).~~~
\end{eqnarray}\end{small}
\end{proposition}
\begin{IEEEproof}
Similar to Proposition 3, $\bar{R}_1^\mathrm{S}$ can be obtained by resolving the integral (\ref{Equ_part2_CR_R1}) over $h^{\max}$ and $g_{n^*_{\mathrm{S}}}^{\max}$. Comparing the distributions of $h^{\max}$ and $g_{n^*_{\mathrm{S}}}^{\max}$ in Proposition 4 to those of $h_{n^*_{\mathrm{P}}}^{\max}$ and $g^{\max}$ in Proposition 3, we can observe that the only difference between them lies in the corresponding upper limits of the summations. In this case, we can calculate $\bar{R}_1^\mathrm{S}$ by replacing the upper limits of $i$ and $j$ in (\ref{Equ_lemma4_rsum_1}) by $NM$ and $K$,~respectively.
\end{IEEEproof}

Finally, by substituting (\ref{Equ_CRA_probability})-(\ref{Equ_lemma5_rsum_1}) into (\ref{Equ_CRA_rsum}), we can obtain an approximated closed-from expression for $\bar{R}_{1}^{\mathrm{MCG}}$.
\begin{remark}
When the primary user UE$_2$ is much closer to the BS than the secondary user UE$_1$, the event that $g^{\max}>h^{\max}$ occurs with a large probability. In this case, the MCG-AS scheme will reduce to the PU-AS scheme and thus $\bar{R}_{1}^{\mathrm{MCG}}\approx\bar{R}_{1}^{\mathrm{P}}$. On the other hand, when UE$_2$ is much further from the BS than UE$_1$, it is more likely that $g^{\max}\leq h^{\max}$. In this case, the MCG-AS scheme degrades to SU-AS, and we have the approximation $\bar{R}_{1}^{\mathrm{MCG}}\approx\bar{R}_{1}^{\mathrm{S}}$.
\end{remark}\vspace{-1em}

\section{Numerical Performance Evaluations}
In this section, the performance of the proposed joint AS algorithms for both F-NOMA and CR-NOMA systems is evaluated by using computer simulations and comparing to the benchmarks. In all simulations, we set $M=K=2$, $\Omega_h=d_1^\alpha$, $\Omega_g=d_2^\alpha$, $d_1$ ($d_2$) is the distance between the BS and UE$_1$ (UE$_2$), the path-loss exponent $\alpha=3$ and the variance of the noise is set as $\sigma^2=-110\mathrm{dBm}$.\vspace{-1em}

\subsection{Numerical Results for F-NOMA Systems}
\begin{figure}
  \centering
  \includegraphics[scale=0.3]{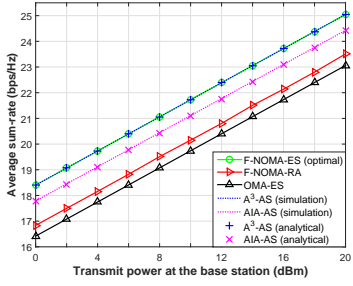}
  \caption{Average sum-rate vs. transmit power in F-NOMA, $N=2$, $d_1=80\mathrm{m}, d_2=200\mathrm{m}, a=0.6, b=0.4$.}
  \label{Fig:Rate_Ps}\vspace{-1em}
\end{figure}

Fig.~\ref{Fig:Rate_Ps} illustrates how the transmit power $P_s$ at the BS affects the system average sum-rate $\bar{R}_{\mathrm{sum}}$. As can be observed from Fig.~\ref{Fig:Rate_Ps}, when $P_s$ increases, $\bar{R}_{\mathrm{sum}}$ increases for all the schemes. Moreover, the performance of the proposed A$^3$-AS and AIA-AS schemes is much better than that of the random AS scheme in F-NOMA scenarios (F-NOMA-RA). Furthermore, the A$^3$-AS scheme can achieve the same performance as that of the optimal ES scheme in F-NOMA scenarios (F-NOMA-ES) but with a much lower computational complexity. We should note that the analytical results match the simulation results for both A$^3$-AS and AIA-AS, which validate our theoretical analysis in Sec. III. It is also worth pointing out that all the NOMA schemes outperform the ES scheme in OMA systems (OMA-ES) over the entire SNR region. For simplicity, we only show the analytical results in the following discussions.

\begin{figure}
  \centering
  \includegraphics[width=0.3\textwidth]{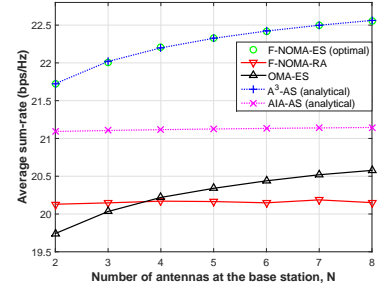}
  \caption{Average sum-rate vs. $N$ in F-NOMA, $d_1=80\mathrm{m}, d_2=200\mathrm{m}, a=0.6, b=0.4, P_s=10\mathrm{dBm}$.}
  \label{Fig:Rate_N}\vspace{-1em}
\end{figure}

Fig.~\ref{Fig:Rate_N} illustrates how the number of antennas $N$ at the BS influences the average sum-rate $\bar{R}_{\mathrm{sum}}$. We can see from this figure that the sum-rates of F-NOMA-RA and AIA-AS keep constant when $N$ increases. For the F-NOMA-RA scheme, this is because it does not properly utilize the multiple-antenna setting but selects one antenna at each node randomly. The reason why AIA-AS keeps constant is because it guarantees the performance of the \textit{weak} user with the poor channel gain $\gamma^w$, but not the \textit{strong} user with the better channel condition $\gamma^s$, which contributes the most to $\bar{R}_{\mathrm{sum}}$. In contrast, the average sum-rate of A$^3$-AS increases along with $N$, and A$^3$-AS achieves the same performance as that of the optimal F-NOMA-ES scheme. Again, all the proposed NOMA schemes outperform the OMA-ES scheme in the entire region.

\begin{figure}
  \centering
  \includegraphics[width=0.3\textwidth]{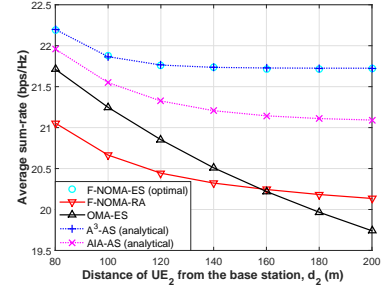}
  \caption{Average sum-rate vs. $d_2$ in F-NOMA, $N=2, d_1=80\mathrm{m}, a=0.6, b=0.4, P_s=10\mathrm{dBm}$.}
  \label{Fig:Rate_d}\vspace{-1em}
\end{figure}

Fig.~\ref{Fig:Rate_d} depicts how the distances between the BS and users influence $\bar{R}_{\mathrm{sum}}$ for the schemes in F-NOMA systems. Take a constant $d_1$ and a variable $d_2$ for example. We can observe that when $d_2$ increases, $\bar{R}_{\mathrm{sum}}$ decreases for all the schemes in F-NOMA systems. We also note that both A$^3$-AS and AIA-AS outperform the F-NOMA-RA and the OMA-ES schemes, and again A$^3$-AS achieves almost the same performance as F-NOMA-ES. There is a crossing between the curves for F-NOMA-RA and OMA-ES. The reason for this is that in OMA-ES, when $d_2$ is much larger than $d_1$, the energy and frequency resources allocated to UE$_2$ are wasted since they contribute very little to $\bar{R}_{\mathrm{sum}}$.

\begin{figure}
  \centering
  \includegraphics[width=0.3\textwidth]{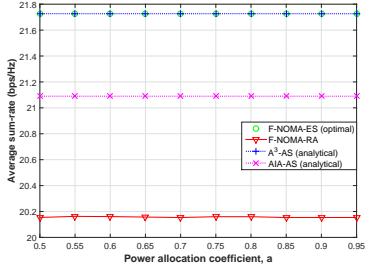}
  \caption{Average sum-rate vs. a in F-NOMA, $N=2, d_1=80\mathrm{m}, d_2=200\mathrm{m}, b=1-a, P_s=10\mathrm{dBm}$.}
  \label{Fig:Rate_b}\vspace{-1em}
\end{figure}

Fig.~\ref{Fig:Rate_b} demonstrates how the power allocation coefficient $b$ affects the $\bar{R}_{\mathrm{sum}}$ for various AS schemes in F-NOMA systems. Interestingly we can see that for all the F-NOMA schemes the average sum-rate keeps almost constant when $b$ increases. The main reason is that ${R}_{\mathrm{sum}}\approx\log(\gamma^s\rho)$ when $\rho\rightarrow\infty$; that is, the sum-rate is not affected by the value of $b$.

\begin{figure}
  \centering
  \includegraphics[width=0.3\textwidth]{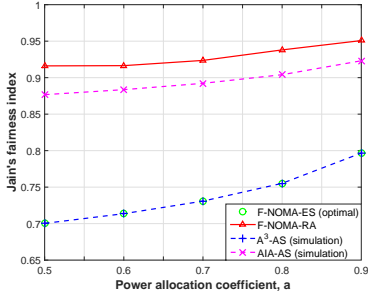}
  \caption{Jain's fairness index vs. b in F-NOMA, $N=4, d_1=80\mathrm{m}, d_2=200\mathrm{m}, b=1-a, P_s=20\mathrm{dBm}$.}
  \label{Fig:fairness}\vspace{-1em}
\end{figure}

Although the system sum-rate performance of A$^3$-AS is shown to be slightly better than that of AIA-AS, regarding the fairness between UE$_1$ and UE$_2$, we can observe in Fig.~\ref{Fig:fairness} that AIA-AS can provide better user fairness than A$^3$-AS. In other words, in practice AIA-AS would be a better trade-off between the system throughput and user~fairness.
\vspace{-1em}

\subsection{Numerical Results for CR-NOMA Systems}
\begin{figure}
  \centering
  \includegraphics[width=0.3\textwidth]{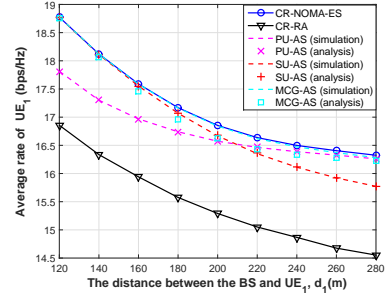}
  \caption{Average rate of UE$_1$ vs. $d_1$ in CR-NOMA, $N=4, d_2=200\mathrm{m}, R_{\mathrm{th}}=5\mathrm{bps/Hz}, P_s=20\mathrm{dBm}$.}
  \label{Fig:CR1_d}\vspace{-1em}
\end{figure}

In the considered two-user CR-NOMA down-link scenario, without loss of generality, we treat UE$_2$ as the PU and UE$_1$ as the SU. Fig.~\ref{Fig:CR1_d} illustrates how the locations of users affect the average rate of UE$_1$. We now fix the location of the primary user UE$_2$ ($d_2=200$m) and place the secondary user UE$_1$ at various locations. As can be observed from Fig.~\ref{Fig:CR1_d}, when $d_1$ increases, the average rate $\bar{R}_1^c$ of UE$_1$  decreases for all the proposed AS schemes in CR-NOMA systems. Interestingly, there is a crossing for the curves of PU-AS and SU-AS around $d_1=d_2=200$m. $\bar{R}_1^c$ achieved by SU-AS is larger than that of PU-AS when $d_1\leq 200$m and the situation is reverse when $d_1>200$m. The reason is that when UE$_1$ is closer to the BS, the channel gain of UE$_1$ contributes more to $\bar{R}_1^c$, but when UE$_2$ is closer to the BS, the channel gain of UE$_2$ is more dominant to $\bar{R}_1^c$ as more power is allocated to the secondary UE. Furthermore, MCG-AS, which takes advantage of PU-AS and SU-AS, can achieve almost the same performance as the optimal ES scheme in CR-NOMA scenarios (CR-NOMA-ES) for all the location settings but with much lower computational complexity, and all the proposed schemes outperform the random selection scheme in CR-NOMA scenarios (CR-NOMA-RA). It is worth pointing out that the analytical results match the simulation results for all the proposed CR-NOMA AS schemes, which validates our theoretical analysis in Sec. IV. For brevity, we only show the analytical results in the following discussions.

\begin{figure}
  \centering
  \includegraphics[width=0.3\textwidth]{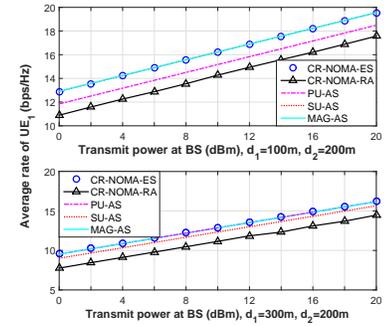}
  \caption{Average rate of UE$_1$ vs. $P_s$ in CR-NOMA, $N=4, R_{\mathrm{th}}=5\mathrm{bps/Hz}.$}
  \label{Fig:CR1_Ps}\vspace{-1em}
\end{figure}

Fig. \ref{Fig:CR1_Ps} illustrates how the transmit power at the BS influences the achievable $\bar{R}_1^c$. We can see that when $P_s$ increases, $\bar{R}_1^c$ increases for all the schemes in CR-NOMA systems. Moreover, all the proposed AS schemes outperform CR-NOMA-RA. In particular, when UE$_1$ is closer to the BS, the performance of SU-AS is better than that of PU-AS. In contrast, when UE$_2$ is closer to the BS, the performances of SU-AS and PU-AS are inverted. Again MCG-AS achieves the near-optimal performance as CR-NOMA-ES in the entire region.

\begin{figure}
  \centering
  \includegraphics[width=0.3\textwidth]{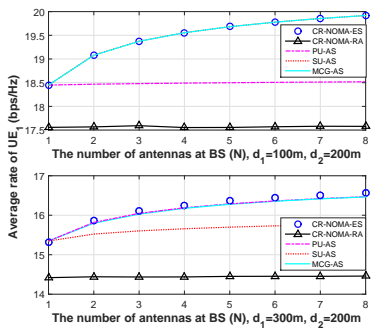}
  \caption{Average rate of UE$_1$ vs. $N$ in CR-NOMA, $P_s=20\mathrm{dBm}, R_{\mathrm{th}}=5\mathrm{bps/Hz}.$}
  \label{Fig:CR1_N}\vspace{-1em}
\end{figure}

Fig. \ref{Fig:CR1_N} illustrates how the number of antennas at the BS affects the achievable $\bar{R}_1^c$. When UE$_1$ is closer to the BS, the curve for SU-AS increases along with $N$. This is because in this case, the channel gain of UE$_1$ contributes more to $\bar{R}_1^c$ and $h_{nk}$ is maximized in SU-AS. In contrast, the curve for PU-AS keeps constant even when $N$ increases. The reason for this is that PU-AS aims at maximizing the channel gain of UE$_2$ instead of UE$_1$. That is, the channel gain of UE$_1$ is selected in a similar way to find out the maximum element from a random row of $\bf H$ and $h_{nk}$ is affected by $M$ but not $N$. Interestingly, when UE$_1$ is farther from the BS than UE$_2$, the aforementioned situation is inverted. Furthermore, all the proposed schemes outperform CR-NOMA-RA, and MCG-AS can achieve a near-optimal performance in the entire~region.

\begin{figure}
  \centering
  \includegraphics[width=0.3\textwidth]{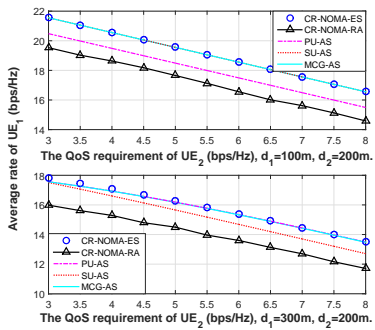}
  \caption{Average rate of UE$_1$ vs. the QoS requirement of UE$_2$ in CR-NOMA, $N=4, P_s=20\mathrm{dBm}$.}
  \label{Fig:CR1_Rth}\vspace{-1em}
\end{figure}

Fig.~\ref{Fig:CR1_Rth} demonstrates how the QoS requirement $R_\mathrm{th}$ affects the achievable $\bar{R}_1^c$. We can see when $R_\mathrm{th}$ increases, the performances decrease for all the AS schemes in CR-NOMA. This is because when $R_\mathrm{th}$ increases, more power is allocated to UE$_2$ to meet the higher QoS requirement and less power can be allocated to UE$_1$. In this case, $\bar{R}_1^c$ would decrease when $R_\mathrm{th}$ increases. Again, all the proposed AS schemes outperform CR-NOMA-RA, and MCG-AS achieves a near-optimal performance in the entire region. Moreover, the performance of PU-AS is better than that of SU-AS when UE$_2$ is closer to the BS, and it is opposite when UE$_1$ is closer to the~BS.

As discussed above, MCG-AS can provide a near-optimal average rate performance for the SU in CR-NOMA networks by taking advantage of the instantaneous channel conditions of all the participants in systems. However, once the network-accessing priority and the near-far relationship of the users in CR-networks are determined, the PU-AS or SU-AS algorithms would be better choices. This is because the computational complexities of PU-AS and SU-AS are further lower than that of MCG-AS, and a near-optimal average rate performance for the secondary UE can be achieved by PU-AU when the primary user is much closer to the BS, and that is achieved by SU-AS when the secondary user is much closer to the BS.\vspace{-1em}

\section{Conclusions}
This paper studied the joint antenna selection (AS) problem for a classical MIMO-NOMA downlink communication scenario. Two computationally efficient AS algorithms, named A$^3$-AS and AIA-AS, were proposed to maximize the system sum-rate of F-NOMA systems without and with the consideration of user fairness, respectively. Meanwhile, a low-complexity AS algorithm (i.e., MCG-AS) and its two simplified versions (i.e., PU-AS and SU-AS) were developed to maximize the rate of the SU in CR-NOMA systems. The asymptotic closed-form expressions for the average sum-rates in F-NOMA systems and average rates of the secondary user in CR-NOMA systems were provided. Numerical simulations demonstrated that in F-NOMA systems, both A$^3$-AS and AIA-AS yield significant performance gains over the random AS scheme and the OMA scheme with exhaustive search-based AS. Furthermore, A$^3$-AS achieves the near-optimal sum-rate performance while AIA-AS provides better user fairness. In CR-NOMA scenarios, MCG-AS can achieve a near-optimal performance with much reduced complexity in the entire region. The complexities of the PU-AS and SU-AS is further lower than that of the MCG-AS scheme, while they only achieve a near-optimal performance when one user is much closer to the base station that the other.\vspace{-1em}
\appendices
\section{Proof of the Proposition \ref{proposition2_A3_rsum}}
When $\rho\rightarrow\infty$, we can attain the asymptotic closed-form expression for the average sum-rate of the A$^3$-AS algorithm as follows:
\begin{small}
\begin{eqnarray}\label{Equ_A3_sum}
\bar{R}_{\mathrm{sum}}^{A^3}\!\!\!\!\!&\approx&\!\!\!\!\!\log\frac{1}{b}+\int_0^\infty\log(1+b\rho_sx)f_{\gamma^s_{A^3}}(x)\mathrm{d}x\overset{(c_2)}{=}\log\frac{1}{b}\nonumber\\
\!\!\!\!\!+&&\!\!\!\!\!\sum_{i=1}^{NM}\sum_{j=1}^{NK}\frac{\mu_{i,NM}\mu_{j,NK}}{\ln2}\left(e^{\frac{i\Omega_h+j\Omega_g}{b\rho}}\mathrm{Ei}\left(-{\frac{i\Omega_h+j\Omega_g}{b\rho}}\right)\right.\nonumber\\
\!\!\!\!\!&&\!\!\!\!\!\left.-e^{\frac{i\Omega_h}{b\rho}}\mathrm{Ei}\left(-{\frac{i\Omega_h}{b\rho}}\right)-e^{\frac{j\Omega_g}{b\rho}}\mathrm{Ei}\left(-{\frac{j\Omega_g}{b\rho}}\right)\right)\nonumber\\
\!\!\!\!\!&\overset{(c_3)}{\approx}&\!\!\!\!\!\log\frac{1}{b}+\sum_{i=1}^{NM}\sum_{j=1}^{NK}\frac{\mu_{i,NM}\mu_{j,NK}}{\ln2}\left(\ln\left(\frac{i\Omega_h+j\Omega_g}{b\rho}\right)\right.\nonumber\\
\!\!\!\!\!&-&\!\!\!\!\!\left.\ln\left(\frac{i\Omega_h}{b\rho}\right)-\ln\left(\frac{j\Omega_g}{b\rho}\right)-C\right)=\log\frac{1}{b}\nonumber\\
\!\!\!\!\!&+&\!\!\!\!\!\sum_{i=1}^{NM}\sum_{j=1}^{NK}\frac{\mu_{i,NM}\mu_{j,NK}}{\ln2}\left(\ln\left(\frac{i\Omega_h+j\Omega_g}{ij\Omega_h\Omega_g}\right)+\ln\left(b\rho\right)-C\right)\nonumber\\
\!\!\!\!\!&\overset{(c_4)}{=}&\!\!\!\!\!\frac{\left(\ln\rho-C\right)}{\ln2}+\sum_{i=1}^{NM}\sum_{j=1}^{NK}\frac{\mu_{i,NM}\mu_{j,NK}}{\ln2}\ln\left(\frac{i\Omega_h+j\Omega_g}{ij\Omega_h\Omega_g}\right).
\end{eqnarray}\end{small}
Specifically, the step $(c_2)$ is obtained according to \cite[Eq. (4.337.2)]{Ref_Book_integrals}, the step $(c_3)$ is obtained by applying the approximation $\mathrm{Ei}(x)\overset{x\rightarrow 0}{\approx}C+\ln|x|$ \cite{Ref_Ei} and the step $(c_4)$ follows the property of the binomial coefficient $\sum_{i=0}^n (-1)^i\binom{n}{i}=0$.\vspace{-1em}

\section{Proof of the Lemma \ref{Lemma1_AIA_gamma_dis}}
In the first stage of AIA-AS, the maximum elements of each row from $\bf H$ and $\bf G$, i.e., $h_n^{\max}$ and $g_n^{\max}$, and their distributions are obtained according to (\ref{Equ_part5_himax_cdf})-(\ref{Equ_part5_gimax_pdf}). In the second stage of AIA-AS, the relatively smaller element $\gamma_{n}^{w}=\min(h_{n}^{\max},g_{n}^{\max})$ in each row for $n \in \mathcal{N}$ is found out. Thus, the CDF of $\gamma_{n}^{w}$ for $x\geq0$ can be calculated as~follows:
\begin{small}
\begin{eqnarray}\label{Equ_lemma1_BW_gamma_w_i_CDF}
F_{\gamma_n^{w}}(x)&=&\mathrm{Pr}\left\{\min(h_{n}^{\max},g_{n}^{\max})<x\right\}\nonumber\\
&=&\mathrm{Pr}\left(h_{n}^{\max}<g_{n}^{\max}<x\right)+\mathrm{Pr}\left(g_{n}^{\max}<h_{n}^{\max}<x\right)\nonumber\\
&=&\int_0^x f_{g_n^{\mathrm{max}}}(y)\int_0^y f_{h_n^{\mathrm{max}}}(z)\mathrm{d}z\mathrm{d}y\nonumber\\
&+&\int_0^x f_{h_n^{\mathrm{max}}}(y)\int_0^y f_{g_n^{\mathrm{max}}}(z)\mathrm{d}z\mathrm{d}y\nonumber\\
&=&1-\sum_{i=1}^{M}\sum_{j=1}^{K}\mu_{i,M}\mu_{j,K}e^{-(i\Omega_h+j\Omega_g)x}.
\end{eqnarray}\end{small}
In the third stage of AIA-AS, $\gamma^w_{\mathrm{AIA}}=\max(\gamma_n^w)$ for $n\in\mathcal{N}$ is selected. In the case that $\gamma^w_{\mathrm{AIA}}$ lies in the $n_{AIA}^*$ row, we first define $\hat{\gamma}^w=\max\limits_{n\neq n^*_{AIA}}(\gamma_n^w)$ and obtain the CDF of $\hat{\gamma}^w$ as~follows:
\begin{eqnarray}\label{Equ_lemma1_BW_gamma_w_uneqnal_CDF}
F_{\hat{\gamma}^w}(x)=[F_{\gamma_n^{w}}(x)]^{N-1}\overset{(c_6)}{=}\sum_{\ell}C_\ell{t_\ell e^{-\xi_\ell x}},
\end{eqnarray}where the step $(c_6)$ is expanded according to the Multinomial theorem. Specifically, $\ell_0+\cdots+\ell_{MK}=N-1$, the multinomial coefficient $C_\ell=\binom{N-1}{\ell_0,\cdots,\ell_{MK}}=\frac{(N-1)!}{\ell_0!\cdots\ell_{MK}!}$, $t_{\ell}=\prod_{\substack{1\leq i\leq M\\1\leq j\leq K}} (-\mu_{i,M}\mu_{j,K})^{\ell_{ij}}$ and $\xi_{\ell}=\sum_{i=1}^M\sum_{j=1}^K(i\Omega_h+j\Omega_j)\ell_{ij}$.

Next we need to obtain the CDF and PDF of $\gamma^s_{\mathrm{AIA}}=\max({h_{n^*_{AIA}}^{\max}},{g_{n^*_{AIA}}^{\max}})$ which lies in the same $n^*_{AIA}$ row as $\gamma^w_{\mathrm{AIA}}$. By applying some algebraic manipulations, we have
\begin{small}
\begin{eqnarray}\label{Equ_lemma1_BW_gamma_s_CDF}
F_{\gamma^s_{\mathrm{AIA}}}(x)&\!\!=\!\!&\mathrm{Pr}\left\{\max(h_{n^*_{AIA}}^{\max},g_{n^*_{AIA}}^{\max})<x, \gamma_{n^*_{AIA}}^w\geq \hat{\gamma}^w \right\}\nonumber\\
&\!\!=\!\!&N\left\{\mathrm{Pr}(\hat{\gamma}^w <g_{n^*_{AIA}}^{\max}<h_{n^*_{AIA}}^{\max}<x)\right.\nonumber\\
&+&\left.\mathrm{Pr}(\hat{\gamma}^w <h_{n^*_{AIA}}^{\max}<g_{n^*_{AIA}}^{\max}<x)\right\},
\end{eqnarray}\end{small}and
\begin{small}\begin{eqnarray}\label{Equ_lemma1_BW_Kappa_n_PDF}
f_{\gamma^s_{\mathrm{AIA}}}(x)&=&N\left(f_{{h_n}^{\max}}(x)\int_0^xf_{{g_n}^{\max}}(y)\int_0^yf_{\hat{\gamma}^w}(z)\mathrm{d}z\mathrm{d}y\right.\nonumber\\
&+&\left.f_{{g_n}^{\max}}(x)\int_0^xf_{{h_n}^{\max}}(y)\int_0^yf_{\hat{\gamma}^w}(z)\mathrm{d}z\mathrm{d}y\right),\nonumber
\end{eqnarray}\end{small}and by applying some algebraic operations, finally $f_{\gamma^s_{\mathrm{AIA}}}(x)$ can be obtained as in (\ref{Equ_part5_AIA_gamma_s_pdf}).\vspace{-2em}

\section{Proof of the Proposition \ref{proposition1_BW_rsum}}
Given the PDF of $\gamma^s_{\mathrm{AIA}}$ \big(i.e., $f_{\gamma^s_{\mathrm{AIA}}}(x)$\big), we can approximate the average sum-rate of AIA-AS \big(i.e., $\bar{R}_{\mathrm{sum}}^{\mathrm{AIA}}$\big) according to (\ref{Equ_part5_FNOMA_rsum_app}) as below:
\begin{eqnarray}\label{Equ_lemma2_BW_rsum}
\bar{R}_{\mathrm{sum}}^{\mathrm{AIA}}&\approx&\log\frac{1}{b}+\int_0^\infty\log(1+b\rho x)f_{\gamma^s_{\mathrm{AIA}}}(x)\mathrm{d}x\\
&=&\log\frac{1}{b}+\sum_{i=1}^M\sum_{j=1}^K\sum_{\ell}\frac{C_\ell t_\ell}{\ln2}\left(T_1+T_2+T_3+T_4\right),\nonumber
\end{eqnarray}where $T_1$ is given by
\begin{eqnarray}
T_1&=&-\int_0^\infty\frac{\xi_{\ell}\zeta_{ij}\ln(1+b\rho x)e^{-j\Omega_gx}}{i\Omega_h\phi_i}\mathrm{d}x\nonumber\\
&\overset{(c_7)}{=}&\frac{\xi_\ell\tilde{\zeta}_{ij}}{\phi_i}e^\frac{j\Omega_g}{b\rho}\mathrm{Ei}\left(-\frac{j\Omega_g}{b\rho}\right)\overset{(c_8)}{\approx}\frac{\xi_\ell\tilde{\zeta}_{ij}}{\phi_i}\chi\left(j\Omega_g\right),
\end{eqnarray}in which $\tilde{\zeta}_{ij}=N\mu_{i,M}\mu_{j,K}$, $\phi_i=i\Omega_h+\xi_\ell$, $\chi(x)=C+\ln|\frac{x}{b\rho}|$, $\mathrm{Ei}(x)$ is the Exponential integral function and $C\approx0.577$ is the Euler's constant. Specifically, the step ($c_7$) is obtained with the help of \cite[Eq. (4.337.2)]{Ref_Book_integrals}, and the step ($c_8$) is obtained by using the approximation of $e^x\approx1$ and $\mathrm{Ei}(x)\approx C+\ln x$ when $x\rightarrow 0$~\cite{Ref_Ei}. Similarly,
\begin{eqnarray}
T_2&=&-\int_0^\infty\frac{\xi_{\ell}\zeta_{ij}\ln(1+b\rho x)e^{-i\Omega_hx}}{j\Omega_g\phi_j}\mathrm{d}x=\frac{\xi_\ell\tilde{\zeta}_{ij}}{\phi_j}\chi(i\Omega_h),\nonumber\\
T_3&=&-\int_0^\infty\frac{\zeta_{ij}\phi_{ij,2}\ln(1+b\rho x)e^{-\phi_{ij,1}x}}{\phi_i\phi_j}\mathrm{d}x=\frac{{\zeta}_{ij}\phi_{ij,2}\chi(\phi_{ij,1})}{\phi_i\phi_j\phi_{ij,1}},\nonumber\\
T_4&=&\int_0^\infty\frac{\zeta_{ij}(i\Omega_h+j\Omega_g)\ln(1+b\rho x)e^{-(i\Omega_h+j\Omega_g)x}}{ij\Omega_h\Omega_g}\mathrm{d}x\nonumber\\
&=&-\tilde{\zeta}_{ij}\chi(i\Omega_h+j\Omega_g),\nonumber
\end{eqnarray}in which, $\phi_j=j\Omega_g+\xi_\ell$, $\phi_{ij,1}=i\Omega_h+j\Omega_g+\xi_\ell$, and $\phi_{ij,2}=i\Omega_h+j\Omega_g+2\xi_\ell$.\vspace{-2em}

\section{Proof of the Proposition \ref{proposition3_CRP_rsum}}
Recall that the achievable rate of the secondary user in (\ref{Equ_part2_CR_R1}) consists of two terms. Here we can calculate the asymptotic average rate of UE$_1$ in PU-AS, which is denoted by $\bar{R}_1^{\mathrm{P}}=\mathbb{E}\left[R_1^\mathrm{P}\right]$, in two parts.

The first part of (\ref{Equ_part2_CR_R1}), i.e., $\bar{R}_{1,1}^{\mathrm{P}}$, can be obtained by calculating the following integral for~$\delta_{n^*,m^*,k^*}=1$:
\begin{small}
\begin{eqnarray}\label{Equ_lemma4_part1}
\bar{R}_{1,1}^{\mathrm{P}}\!\!\!\!\!&=&\!\!\!\!\!\int_0^\infty\int_0^x \log\left(\frac{\rho x}{\varepsilon+1}\right)f_{g^{\max}}(y)f_{{h_n}^{\max}}(x)\mathrm{d}y\mathrm{d}x\nonumber\\
\!\!\!\!\!&=&\!\!\!\!\!\sum_{i=1}^M\sum_{j=1}^{NK}\frac{i\Omega_h\mu_{i,M}\mu_{j,NK}}{\ln 2}\left\{\int_0^\infty\ln\left(\frac{\rho x}{\varepsilon+1}\right)e^{-i\Omega_hx}\mathrm{d}x\right.\nonumber\\
\!\!\!\!\!&-&\!\!\!\!\!\left.\int_0^\infty\ln\left(\frac{\rho x}{\varepsilon+1}\right)e^{-(i\Omega_h+j\Omega_g)x}\mathrm{d}x\right\}\nonumber\\
\!\!\!\!\!&\overset{y=\rho x/(\varepsilon+1)}{=}&\!\!\!\!\!\sum_{i=1}^M\sum_{j=1}^{NK}\frac{i(\varepsilon+1)\Omega_h\mu_{i,M}\mu_{j,NK}}{\rho\ln 2}\left\{\int_0^\infty\ln ye^{-\frac{(\varepsilon+1)i\Omega_h y}{\rho}}\mathrm{d}y\right.\nonumber\\
\!\!\!\!\!&-&\!\!\!\!\!\left.\int_0^\infty\ln ye^{-\frac{(\varepsilon+1)(i\Omega_h+j\Omega_g)y}{\rho}}\mathrm{d}y\right\}\nonumber\\
\!\!\!\!\!&\overset{(c_9)}{=}&\!\!\!\!\!\sum_{i=1}^M\sum_{j=1}^{NK}\frac{\mu_{i,M}\mu_{j,NK}}{\ln2}\left\{-C-\ln\left(\frac{i(\varepsilon+1)\Omega_h}{\rho}\right)\right.\nonumber\\
\!\!\!\!\!&+&\!\!\!\!\!\left.\frac{i\Omega_hC}{i\Omega_h+j\Omega_g}+\frac{i\Omega_h\ln\left(\frac{(\varepsilon+1)(i\Omega_h+j\Omega_g)}{\rho}\right)}{i\Omega_h+j\Omega_g}\right\}\nonumber\\
\!\!\!\!\!&=&\!\!\!\!\!\sum_{i=1}^M\sum_{j=1}^{NK}\frac{\mu_{i,M}\mu_{j,NK}}{\ln2}\left\{-\ln\left(i\Omega_h\right)\right.\nonumber\\
\!\!\!\!\!&+&\!\!\!\!\!\left.\frac{j\Omega_g\left(\ln\left(\frac{\rho}{\varepsilon+1}\right)-C\right)+i\Omega_h\ln\left(i\Omega_h+j\Omega_g\right)}{i\Omega_h+j\Omega_g}\right\}.
\end{eqnarray}\end{small}Specifically, the step $(c_9)$ is obtained with the help of \cite[Eq. (4.331.1)]{Ref_Book_integrals}.

The second part of (\ref{Equ_part2_CR_R1}), i.e., $\bar{R}_{1,2}^{\mathrm{P}}$, can be obtained by calculating the following integral for $\delta_{n^*,m^*,k^*} = 0$:
\begin{small}
\begin{eqnarray}\label{Equ_lemma4_part2}
\bar{R}_{1,2}^{\mathrm{P}}\!\!\!&=&\!\!\!\int_0^\infty\int_x^\infty\log\left(\frac{\rho xy}{\varepsilon x+y}\right)f_{g^{\max}}(y)\mathrm{d}yf_{h_{n}^{\max}}(x)\mathrm{d}x\nonumber\\
\!\!\!&=&\!\!\!\underbrace{\int_0^\infty\int_x^\infty\log\left({\rho x}\right)f_{g^{\max}}(y)\mathrm{d}yf_{h_{n}^{\max}}(x)\mathrm{d}x}_{T_5}\nonumber\\
\!\!\!&+&\!\!\!\underbrace{\int_0^\infty\int_x^\infty\log\left(y\right)f_{g^{\max}}(y)\mathrm{d}yf_{h_{n}^{\max}}(x)\mathrm{d}x}_{T_6}\nonumber\\
\!\!\!&-&\!\!\!\underbrace{\int_0^\infty\int_x^\infty\log\left(\varepsilon x+y\right)f_{g^{\max}}(y)\mathrm{d}yf_{h_{n}^{\max}}(x)}_{T_7}\mathrm{d}x,~~~
\end{eqnarray}\end{small}
in which,
\begin{small}
\begin{eqnarray}\label{Equ_lemma4_T5}
T_5\!\!\!\!\!\!\!&=&\!\!\!\!\!\!\!\sum_{i=1}^{M}\sum_{j=1}^{NK}\frac{ij\Omega_h\Omega_g\mu_{i,M}\mu_{j,NK}\int_0^\infty\!\!\int_x^\infty\ln\left({\rho x}\right)e^{-j\Omega_gy}\mathrm{d}ye^{-i\Omega_hx}\mathrm{d}x}{\ln2}\nonumber\\
\!\!\!\!\!\!\!&=&\!\!\!\!\!\!\!\sum_{i=1}^{M}\sum_{j=1}^{NK}\frac{i\Omega_h\mu_{i,M}\mu_{j,NK}}{\ln2}\int_0^\infty\ln\left({\rho x}\right)e^{-(i\Omega_h+j\Omega_g)x}\mathrm{d}x\nonumber\\
\!\!\!\!\!\!\!&\overset{(c_{10})}{=}&\!\!\!\!\!\!\! \sum_{i=1}^{M}\sum_{j=1}^{NK}\frac{\mu_{i,M}\mu_{j,NK}\left(-i\Omega_h\left(C+\ln(i\Omega_h+j\Omega_g)-\ln\rho\right)\right)}{\ln2\left(i\Omega_h+j\Omega_g\right)},~
\end{eqnarray}\end{small}where the step $(c_{10})$ is obtained with the help of the substitution $t=\rho x$ and that of \cite[Eq. (4.331.1)]{Ref_Book_integrals}. Similarly,
\begin{small}
\begin{eqnarray}\label{Equ_lemma4_T6}
T_6\!\!\!\!\!&=&\!\!\!\!\!\sum_{i=1}^{M}\sum_{j=1}^{NK}\frac{ij\Omega_h\Omega_g\mu_{i,M}\mu_{j,NK}}{\ln2}\int_0^\infty\int_x^\infty\ln ye^{-j\Omega_gy}\mathrm{d}ye^{-i\Omega_hx}\mathrm{d}x\nonumber\\
\!\!\!\!\!&\overset{(c_{11})}{=}&\!\!\!\!\!\sum_{i=1}^{M}\sum_{j=1}^{NK}\frac{i\Omega_h\mu_{i,M}\mu_{j,NK}}{\ln2}\int_0^\infty \!\!\!\left(\ln x e^{-j\Omega_gx}\right.\nonumber\\
\!\!\!\!\!&-&\!\!\!\!\!\left.\mathrm{Ei}\left(-j\Omega_gx\right)\right)e^{-i\Omega_hx}\mathrm{d}x\nonumber\\
\!\!\!\!\!&\overset{(c_{12})}{=}&\!\!\!\!\!\sum_{i=1}^{M}\sum_{j=1}^{NK}\frac{\mu_{i,M}\mu_{j,NK}}{\ln2}\left(\ln\left(i\Omega_h+j\Omega_g\right)\right.\nonumber\\
\!\!\!\!\!&-&\!\!\!\!\!\left.\ln\left(j\Omega_g\right)-\frac{i\Omega_h\left(C+\ln\left(i\Omega_h+j\Omega_g\right)\right)}{i\Omega_h+j\Omega_g}\right),
\end{eqnarray}\end{small}where the step $(c_{11})$ is calculated with the substitution $t=y/x$ and \cite[Eq. (4.331.2)]{Ref_Book_integrals}, and the step $(c_{12})$ is with \cite[Eq. (2.5.3.1)]{Ref_Book_integrals2}. For the $T_7$, we have
\begin{small}
\begin{eqnarray}\label{Equ_lemma4_T7}
T_7\!\!\!\!\!&=&\!\!\!\!\!\sum_{i=1}^{M}\sum_{j=1}^{NK}\frac{ij\Omega_h\Omega_g\mu_{i,M}\mu_{j,NK}}{\ln2}\int_0^\infty\int_x^\infty\ln\left(\varepsilon x+y\right)\nonumber\\
\!\!\!\!\!&&\!\!\!\!\!e^{-j\Omega_gy}\mathrm{d}ye^{-i\Omega_hx}\mathrm{d}x\nonumber\\
\!\!\!\!\!&\overset{t=y-x}{=}&\!\!\!\!\!\sum_{i=1}^{M}\sum_{j=1}^{NK}\frac{ij\Omega_h\Omega_g\mu_{i,M}\mu_{j,NK}}{\ln2}\int_0^\infty\int_0^\infty\ln\left((\varepsilon+1) x+t\right)\nonumber\\
\!\!\!\!\!&&\!\!\!\!\!e^{-j\Omega_gt}\mathrm{d}te^{-(i\Omega_h+j\Omega_g)x}\mathrm{d}x\nonumber\\
\!\!\!\!\!&\overset{(c_{13})}{=}&\!\!\!\!\!\sum_{i=1}^{M}\sum_{j=1}^{NK}\frac{i\Omega_h\mu_{i,M}\mu_{j,NK}}{\ln2}\int_0^\infty \left(\ln(\varepsilon+1)+\ln x\right.\nonumber\\
\!\!\!\!\!&-&\!\!\!\!\!\left.\mathrm{Ei}\left[-j\Omega_g(\varepsilon+1)x\right]e^{j\Omega_g(\varepsilon+1)x}\right)e^{-(i\Omega_h+j\Omega_g)x}\mathrm{d}x\nonumber\\
\!\!\!\!\!&\overset{(c_{14})}{=}&\!\!\!\!\!\sum_{i=1}^{M}\sum_{j=1}^{NK}\frac{\mu_{i,M}\mu_{j,NK}}{\ln2}\left(\frac{i\Omega_h\left(\ln(\varepsilon+1)-C-\ln(i\Omega_h+j\Omega_g)\right)}{i\Omega_h+j\Omega_g}\right.\nonumber\\
\!\!\!\!\!&+&\!\!\!\!\!\left.\frac{i\Omega_h\left(\ln(i\Omega_h+j\Omega_g)-\ln\left((\varepsilon+1)j\Omega_g\right)\right)}{i\Omega_h-\varepsilon j\Omega_g}\right)
\end{eqnarray}\end{small}
where the step $(c_{13})$ is with \cite[Eq. (4.337.1)]{Ref_Book_integrals} and the step $(c_{14})$ is with \cite[Eq. (4.331.1)]{Ref_Book_integrals} and  \cite[Eq. (2.5.3.1)]{Ref_Book_integrals2}.

In this case, we can obtain $\bar{R}_{1,2}^{\mathrm{P}}$ by summing up (\ref{Equ_lemma4_T5})-(\ref{Equ_lemma4_T7}), and obtain $\bar{R}_{1}^{\mathrm{P}}$ by summing up (\ref{Equ_lemma4_part1}) and (\ref{Equ_lemma4_part2}). After some manipulations, $\bar{R}_{1}^{\mathrm{P}}$ can be obtained as in (\ref{Equ_lemma4_rsum_1}).\vspace{-1em}

\ifCLASSOPTIONcaptionsoff
  \newpage
\fi

\vspace{-2em}

\end{document}